\documentclass[aps,prb,showpacs,twocolumn
]{revtex4-1}
\usepackage{graphicx,amssymb,amsfonts,amsmath}
\begin{document}
\title{Correlation functions in the prethermalized regime after a quantum quench of a spin-chain}
\author{Aditi Mitra}
\affiliation{Department of Physics, New York University,
4 Washington Place, New York, New York 10003, USA}

\date{\today}

\pacs{75.10.Jm,05.70.Ln,67.85.-d,71.10.Pm}


\begin{abstract}
Results are presented for a two-point correlation function
of a spin-chain after a quantum quench for an intermediate time regime where
inelastic effects are weak. A Callan-Symanzik like equation for the
correlation function is explicitly constructed which is used to show the appearance of 
three distinct scaling regimes. One is for spatial separations within a light-cone, the
second is for spatial separations on the light-cone, and the third is for
spatial separations outside the light-cone. In these three regimes, the correlation function is
found to decay with power-laws with nonequilibrium exponents that differ from 
those in equilibrium, as well as from those obtained from quenches in a quadratic Luttinger liquid theory.
A detailed discussion is presented on how the existence of scaling depends on the properties of the
initial state before the quench.
\end{abstract}

\maketitle

\section{Introduction} \label{intro}

Motivated by experiments in cold atomic gases~\cite{Bloch08} and ultra-fast spectroscopy of strongly correlated 
materials~\cite{Fausti11,Smallwood12},
the nonequilibrium dynamics of interacting quantum systems has become a topic at the fore-front of research.
In this context, dynamics arising due to a quantum quench, where a system is prepared in 
the ground state of an initial Hamiltonian $H_i$, and then
time-evolved with respect to a final Hamiltonian $H_f$, is of particular interest because of its potential
for addressing several fundamental questions~\cite{Polkovnikovrev}. Some of these are, 
the mechanisms and time-scales for thermalization~\cite{Rigol09b,Santos10,Biroli10,Mitra11,Mitra12a,Eisert12,Sirker13}, 
the possibility of an intermediate
time prethermalized regime~\cite{Berges04,Kehrein08,Kollar08,Sabio10}, dynamical phase transitions associated with
non-analytic behavior during the time 
evolution~\cite{Heyl13,Mitra12b}, dynamics of integrable 
models~\cite{Cazalilla06,Barthel08,Lancaster10,Caux11,Rentrop12,Essler12,Calabrese11,Andrei12} and the possibility of describing their 
steady-state in terms of a generalized Gibbs ensemble (GGE)~\cite{Rigol07}. Yet another important question, which is
related to the topic of this paper, is
the appearance of universal behavior in the dynamics
with the possibility of capturing such
a behavior in a renormalization group (RG) approach, even though the system is out of equilibrium.

An RG approach has been actively used to study nonequilibrium time-evolution after a quench in classical 
field theories~\cite{Janssen89,Gambassi05}. The aim of this paper is to develop an RG approach to study nonequilibrium
time-evolution of correlation functions in interacting quantum field theories. So far the establishment of universal scaling functions 
in quantum systems that are out of equilibrium either due to a sudden quench
or a ``slow'' quench that involves changing parameters in a prescribed time-dependent way, has been mainly explored 
for exactly solvable or mean field theories~\cite{Chandran12,Biroli12}, or in numerical studies of interacting
field theories~\cite{deGrandi11,Huse12,Torre12}. 
In the present paper on the other hand we present an analytical RG approach to study how a correlation function evolves after 
a quantum quench in an interacting field theory. In doing so a Callan-Symanzik (CS) like 
equation for a two-point correlation function is derived which
is used to explicitly show under what conditions scaling holds out of equilibrium, and is 
used to identify intrinsically nonequilibrium
scaling regimes with new exponents.

We study a quantum quench in 
a generic one-dimensional spin-chain $H=\sum_i\left[S^x_iS^x_{i+1}+
S^y_i S^y_{i+1}+\Delta S^z_i S^z_{i+1}\right] + n.n.n.$ where $n.n.n.$ denote additional next-nearest-neighbor 
couplings. We study the dynamics in a continuum field theory described by the quantum sine-Gordon model
with the cosine potential representing the underlying commensurate lattice or periodic potential~\cite{Giamarchibook}. 
In particular we study the
time-evolution of the two-point correlation function of the staggered spin component
$R(r,T_{m})=(-1)^r\langle \psi_i|e^{i H_f T_{m}} S^z_0S^z_re^{-i H_f T_{m}}|\psi_i\rangle$ where
$|\psi_i\rangle$ is the ground state of the Hamiltonian $H_i$ before the quench, $H_f$ is the Hamiltonian after the quench, 
$T_{m}$ is the time after the quench and $r$ is the spatial separation between spin operators. We find that when the quench involves the
sudden switching on of the cosine potential, then in the vicinity of the critical point where the cosine potential is a
marginal perturbation in equilibrium~\cite{Singh89,Affleck98,Barzykin99}, out of equilibrium three distinct scaling regimes appear
for macroscopic distance and time scales $r,T_m\gg 1$ (we have set the sound velocity $u=1$, and the distances and times are measured in
units of an ultra-violet (UV) cutoff). 
One of the scaling regimes is for
spatial separations outside the light-cone ($r\gg 2T_m$) 
where we find $R(r,T_m, r \gg 2 T_m)\sim \frac{\sqrt{\ln{T_m}}}{r}$. The second scaling regime is for spatial separations inside the light-cone
$r\ll 2 T_m$ where we find $R(r,T_m, r \ll 2 T_m)\sim \frac{\sqrt{\ln{r}}}{r}$. The third scaling regime is for 
spatial separations on the light-cone $r=2T_m$ where we find $R(r,T_m, r= 2 T_m)\sim \frac{\ln{r}}{r}$.

For more complicated quenches which involve not only a sudden switching on of the cosine potential, but also a change
in the Luttinger interaction parameter from $K_0 \rightarrow K$, we find that the scaling within the light-cone survives,
where the correlation function is found to be $R(r,T_m,r\ll 2 T_m)\sim \frac{\left(\ln{r}\right)^\theta}{r}$ where $\theta$
is a universal number that approaches $1/2$ as $K_0\rightarrow K$. In addition we also find that the
existence of scaling in the other two regimes, one being outside the light-cone ($r\gg 2 T_m$), and the second being on the
light-cone $r=2T_m$ depends on the initial wave-function. In particular if the initial Luttinger parameter $K_0$ is such that
the cosine potential is a relevant or marginal perturbation, scaling holds on the light-cone. Whereas scaling 
on the light-cone is violated when the cosine potential is an irrelevant perturbation
for the initial state. 

This paper is organized as follows. In Section~\ref{results} we introduce the model, establish notation and also briefly summarize the results.
The rest of the paper goes into more details of how these results are obtained. 
In Section~\ref{quadratic} we briefly present results for an interaction
quench in the quadratic theory (the Luttinger liquid) in the language of Keldysh Green's functions. 
These results will be useful for later sections when we perform perturbation theory in the cosine
potential. In Section~\ref{beta}, we perform perturbation theory in the cosine potential and derive the $\beta$-function to 
two loop.  
In Section~\ref{pertR} we present results for the correlation function within perturbation theory to leading order in
the cosine potential. These results then set the stage for doing renormalization improved perturbation theory which will be
explicitly carried out in Section~\ref{CSder} where a CS like differential equation for the correlation function is
derived. Results of the solution of the CS equation are presented in Section~\ref{lattice} for the case where only the
cosine or lattice potential is suddenly switched on, while results for the correlation function for a simultaneous lattice and interaction
quench are presented in Section~\ref{latticeint}. Finally in Section~\ref{summary} we present our conclusions.
 
\section{Model and a brief discussion of results} \label{results}

We study a quantum quench in 
a generic one-dimensional (1D) spin-chain 
\begin{eqnarray}
H=\sum_i\left[S^x_iS^x_{i+1}+
S^y_i S^y_{i+1}+\Delta S^z_i S^z_{i+1}\right] + n.n.n.
\end{eqnarray} 
where $n.n.n.$ denote additional next-nearest-neighbor couplings. 
If the $n.n.n.$ couplings are weak in comparison to the $n.n.$ couplings, the
spin-chain has two phases, a gapless phase with linearly dispersing spin-waves at long wavelengths, and
a gapped antiferromagnetic Ising phase. 
In equilibrium and zero temperature
the properties of the spin-chain in its gapless phase are captured very well by
a continuum theory that retains only the relevant operators, namely the Luttinger liquid~\cite{Giamarchibook}. 
In contrast, the effect of irrelevant operators can be important both at finite
temperature~\cite{Sachdev94,Castella95,Zotos99,Rosch00,Altshuler06,Sirker09} as well as out of equilibrium following a 
quench~\cite{Mitra11,Lancaster11}. In this paper we study the dynamics of the spin-chain 
in a continuum theory by retaining the effect of the leading irrelevant operator.
For the spin-chain in its gapless phase, the leading irrelevant operator
is a commensurate periodic 
or lattice potential which gives rise to Umpklapp or
back-scattering. In this paper we study the effect of this
term on the dynamics, we expect its 
effect will dominate over those of 
other irrelevant terms such as band curvature. 

Specifically 
we study a quench where initially the system is in the ground-state of
a Luttinger liquid
\begin{eqnarray}
&&H_i = \frac{u_0}{2\pi}\int dx
\biggl\{K_0\left[\pi \Pi(x)\right]^2 + \frac{1}{K_0}\left[\partial_x \phi(x)\right]^2
\biggr\}
\label{Hidefa}
\end{eqnarray}
$-\partial_x\phi/\pi$ represents the density, $\Pi$ is the
variable canonically conjugate to $\phi$, $K_0$ is the dimensionless interaction
parameter, and $u_0$ is the velocity of the sound modes.
The system is driven out of equilibrium via
an interaction quench at $t$=$0$ from $K_0 \rightarrow K$, with the leading irrelevant operator corresponding to a commensurate lattice 
or periodic potential 
$V_{sg}$ also switched on suddenly, at the same time as the interaction quench.
This triggers non-trivial time-evolution
from $t >0$ due to the quantum sine-Gordon model,
\begin{eqnarray}
H_f=H_{f0} + V_{sg}
\end{eqnarray} 
where 
\begin{eqnarray}
H_{f0}= \frac{u}{2\pi}\int dx
\biggl\{K\left[\pi \Pi(x)\right]^2 + \frac{1}{K}\left[\partial_x \phi(x)\right]^2
\biggr\}\\
V_{sg} = -\frac{gu}{\alpha^2} \int dx \cos(\gamma \phi)
\end{eqnarray}
Above
\begin{eqnarray}
\Lambda =\frac{u}{\alpha}
\end{eqnarray}
is a short-distance UV cut-off, $g$ is the strength of the commensurate periodic potential. Note that
both for the n.n. spin-chain as well as the n.n.n spin-chain, the low energy theory is represented by $H_f$ with  
$\gamma=4$. However the  precise values of the Luttinger parameter $K$ and the
strength of the cosine potential $g$ depends on the microscopic details. For example, 
the point $K=1,g=0$ corresponds to the
exactly solvable n.n. $XX$ chain. Switching on $n.n.n$ interactions can result in parameters where $K \neq 1$
but $g=0$~\cite{Haldane82}. 

The versatility of $H_f$ is that it equally well applies to interacting bosons 
in a commensurate periodic potential~\cite{Cazalillarev11}. For this case, 
$\gamma=2$, while the point $K=1$ corresponds to n.n. hard-core bosons or the Tonks-Girardeau gas.  
In this paper, in all our analytic results, we keep $\gamma$ general so that the obtained results may be applied 
both to the spin-chain as well as to bosons in a commensurate periodic potential.

\begin{figure}
\centering
\includegraphics[width=6cm]{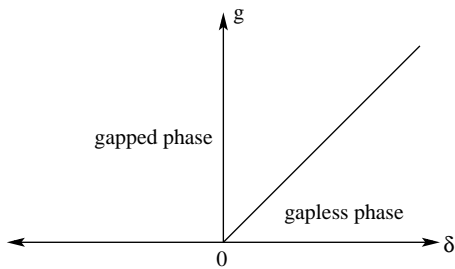}
\caption{Ground state phase diagram of the quantum sine-Gordon model. A critical line separates a gapless phase where the
cosine term is irrelevant from a gapped phase where the cosine term is relevant. The critical line is 
located at $\delta = 2\pi g$, where $\delta = K_{eq}-2$, with $K_{eq}=\frac{\gamma^2K}{4}$.
}
\label{fig1}
\end{figure}

The ground state phase diagram of the quantum sine-Gordon model is shown in Fig.~\ref{fig1}.
A critical line defined by $\delta=2\pi g$ where,
$\delta = K_{eq}-2$, with $K_{eq}=\frac{\gamma^2K}{4}$, 
separates a gapped phase where $V_{sg}$ is a relevant perturbation, from a gapless phase, where 
$V_{sg}$ is an irrelevant perturbation. For the spin-chain the
gapped phase corresponds to the Ising phase, while for interacting
bosons in a lattice, the gapped phase is the Mott insulator.

In this paper we study the time-evolution
of the equal time two-point correlation function 
of the staggered spin component $R(r,T_m)=(-1)^r\langle S^z_0(T_m)S^z_r(T_m)\rangle$ where $T_m$ is the time
after the quench. In the continuum, this correlator is given by~\cite{comm1}
\begin{eqnarray}
&&R(x_1T_m,x_2T_m) = 4\langle\cos\left(\frac{\gamma \phi(x_1T_m)}{2}\right)\cos\left(\frac{\gamma\phi(x_2T_m)}{2}\right) \rangle 
\nonumber\label{Rdef}\\
\end{eqnarray}
In equilibrium, and in the gapless phase, but in the vicinity of the critical line where
$V_{sg}$ is a marginal perturbation, 
logarithmic corrections arise. In particular $R$ near the equilibrium critical point 
behaves as follows~\cite{Singh89,Affleck98,Barzykin99},
\begin{eqnarray}
R_{eq}(r) = \frac{\sqrt{\ln{r}}}{r} + {\cal O}\left(\frac{1}{r^2}\right)\label{req}
\end{eqnarray}
where $r$ is the magnitude of the spatial separation $x_1-x_2$.
The aim of this paper is to determine how the correlator $R$ evolves after 
a quantum quench. We will study $R$ in the regime where $V_{sg}$ is irrelevant or marginally irrelevant, where the
meaning of these terms in a nonequilibrium situation will be clarified below.
 
We now briefly outline how $R$ is calculated for the nonequilibrium problem.
Denoting $\hat{O}(xt) = 2\cos\left(\frac{\gamma\phi(xt)}{2}\right)$,
$R$ may be written as a Keldysh path-integral representing the 
time-evolution from the initial pure state $|\psi_i\rangle$ (hence an initial density matrix 
$\rho =|\psi_i\rangle\langle\psi_i| $) corresponding to the ground state of $H_i$,
\begin{eqnarray}
R(x_1T_m,x_2T_m) =Tr\left[\rho(T_m)O(x_10)O(x_20)\right]
\end{eqnarray} 
which may be written as
\begin{eqnarray}
&&R(x_1T_m,x_2T_m)=\nonumber \\
&&Tr\left[e^{-iH_f T_m}|\psi_i\rangle\langle\psi_i|e^{i H_f T_m}\hat{O}(x_1)\hat{O}(x_2)\right]\nonumber \\
&&=\int {\cal D}\left[\phi_{cl},\phi_q\right] e^{i \left(S_0 + S_{sg}\right)}\hat{O}_I(x_1T_m)\hat{O}_I(x_2T_m)
\end{eqnarray}
where $\hat{O}_I$ is the operator in the interaction representation of $H_{f0}$,
$S_0$ describes the nonequilibrium Luttinger liquid ($g$=$0$),
\begin{eqnarray}
&&S_0 = \nonumber \\
&&\frac{1}{2}\int_{-\infty}^{\infty} dx_1 \int_{-\infty}^{\infty} dx_2\int_0^{T_m} dt_1\int_0^{T_m} dt_2
\begin{pmatrix} \phi_{cl}(1) & \phi_q(1)\end{pmatrix}\nonumber \\
&&\begin{pmatrix} 0&&G_A^{-1}(1,2)\\
G_R^{-1}(1,2) && -\left[G_R^{-1} G_K G^{-1}_A\right](1,2)
\end{pmatrix}\begin{pmatrix}
\phi_{cl}(2)\\
\phi_q(2)
\end{pmatrix}\label{S0a}
\end{eqnarray}
and
\begin{eqnarray}
&&S_{sg}=\nonumber \\
&&\!\!\frac{gu}{\alpha^2}\int_{-\infty}^{\infty}dx_1 \int_0^{T_m} dt_1\!\!\left[\cos\{\gamma\phi_-(1)\}\
-\cos\{\gamma\phi_{+}(1)\}\right]
\end{eqnarray} 
Above
$1(2)$=$(x_{1(2)},t_{1(2)})$, 
$\phi_{cl,q}$=$\frac{\phi_-\pm \phi_+}{\sqrt{2}}$
with $-/+$ representing fields that are time/anti-time ordered on the Keldysh contour~\cite{Kamenevbook}.
\begin{figure}
\centering
\includegraphics[totalheight=4cm]{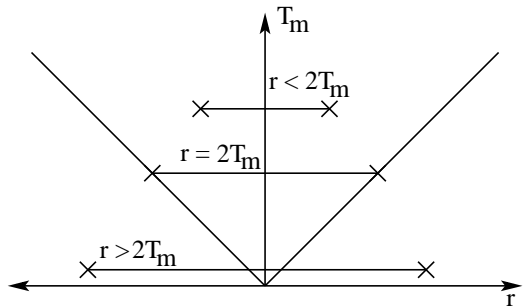}
\caption{
The equal-time correlator shows distinctly different behavior 
for the following three cases: spatial separations outside the light-cone ($r > 2T_m$), spatial separations on the
light-cone ($r=2T_m$), and spatial separations within the light-cone ($r<2T_m$). The present paper gives results for correlation
functions in these three regimes under the additional constraint that $T_m < 1/\eta$ where
$\eta$ is an inelastic scattering rate.}
\label{fig2}
\end{figure}

We derive a CS-like differential equation for 
$R$ by splitting the fields 
$\phi$ into slow and fast fields $\phi = \phi_< + \phi_>$ where the fast fields have
a large weight at short wavelengths, and therefore oscillate rapidly in time. We integrate out the fast fields, 
and rescale the cut-off,
position and time. Such a procedure within the real-time Keldysh
approach has been employed for the quantum sine-Gordon model both for 
steady-state~\cite{Mitra11,Mitra12a,Torre12} and transient behavior~\cite{Mitra12b}, 
where in each case the $\beta$ function was derived.  
Here we generalize this approach to the study of a two-point correlation function by performing a 
microscopic derivation of a CS like differential equation for the correlation function. 
This approach reveals the conditions under which
scaling holds after a quantum quench, and identifies different scaling regimes.
So far such a treatment has only been employed for quenches in classical field theories
where intermediate time nonequilibrium scaling regimes with new exponents have been 
identified~\cite{Janssen89,Gambassi05}. Here we will show that similar new nonequilibrium scaling regimes
can arise for quantum quenches of 1D systems. While interaction quenches in Luttinger liquids and related quadratic theories have been
studied extensively~\cite{Cazalilla06,Iucci09,Kennes10,
Lancaster10,Perfetto06,Dora11,Perfetto11,Rentrop12}, and predict new nonequilibrium exponents as well, in this paper we show
that these exponents are further modified by the presence of the commensurate periodic potential $V_{sg}$. For example
when $V_{sg}$ is marginal, as briefly
stated in the introduction, $V_{sg}$ gives rise to logarithmic corrections. 

In a global quench like the one we study, the system is translationally invariant
in space, so that $R(x_1T_m,x_2T_m)= R(r=x_1-x_2,T_m)$.
We define the following three exponents which play an important role in the dynamics,
\begin{eqnarray}
K_{eq}=\frac{\gamma^2 K}{4};
K_{neq}=\frac{\gamma^2}{8}K_0\left(1+\frac{K^2}{K_0^2}\right)\nonumber\\
K_{tr}=\frac{\gamma^2}{8}K_0\left(1-\frac{K^2}{K_0^2}\right)\label{Kexp}
\end{eqnarray}
$K_{eq}$ governs the power-law decay of the correlator $R$ in the ground state of
$H_{f0}$ ({\sl i.e.}, the Luttinger liquid with interaction parameter $K_0$ and $V_{sg}=0$),
$K_{neq}$ determines the power-law decay of $R$
at long times after an interaction quench in the Luttinger
liquid ($V_{sg}=0$)~\cite{Iucci09,Mitra11}, $K_{tr}$ determines the crossover from a short-time behavior determined 
primarily by the initial Luttinger parameter $K_{neq}+K_{tr}=\frac{\gamma^2K_0}{4}$, to the long time behavior determined by
$K_{neq}$~\cite{Mitra12b}. 

One of the results of our study is that the dynamics of equal-time correlation functions after a quench has 
qualitatively different features in the three regimes shown
in Fig.~\ref{fig2}. One is the region outside the light-cone where the spatial separation $r$ is much larger than the time after the quench
($r\gg 2 T_m$ setting the velocity $u=1$), here the behavior of the correlator is primarily determined by the initial wave-function. 
The second is an intrinsically nonequilibrium regime where the separation $r$ lies
on the light-cone ($r=2T_m$) and where we will identify 
universal behavior of the correlator $R$ with new exponents. 
The third is a nonequilibrium steady-state regime where the spatial separation lies within
the light-cone ($r \ll 2T_m$). Here $R$ is independent of time and shows scaling behavior in
position with new nonequilibrium exponents which differ from those
in the first two regimes just discussed. This qualitative change in the behavior of the correlators at $r=2T_m$
is known as the ``horizon effect''~\cite{Calabrese06} where at time $T_m=r/2$ left and right moving excitations
originating from the same spatial region reach the two local observables, thus maximally entangling them. 
 
The presence of these three regimes is already apparent in the behavior of $R$ for an interaction quench in
a Luttinger liquid ($V_{sg}=0$). 
Here (writing $r,T_m$ in units of $\Lambda$) for 
$r,T_m \gg 1$ and far from the light-cone ($|r \pm 2T_m| \gg 1$) 
one finds~\cite{Iucci09,Mitra12b}, 
\begin{eqnarray}
&&R^{(0)}(r,T_m,g=0) = \nonumber\\
&&\left[\left(\frac{1}{r^2}\right)^{\frac{\gamma^2K_0}{4}}\left(\lvert \frac{r^2 - (2T_m)^2}{r^2(2T_m)^2}\rvert\right)^{-K_{tr}}
\right]^{1/4}\label{R0ll}
\end{eqnarray}
$R$ shows the ``horizon-effect'' where outside the light-cone $r\gg 2T_m$, the correlator depends only
on the initial wave-function (and hence the initial Luttinger parameter $K_0$) albeit with a time-dependent prefactor 
$T_m^{K_{tr}/2}$. However within the light-cone $r\ll 2 T_m$,  the correlator reaches a steady-state
characterized by the nonequilibrium exponent $K_{neq}$. Moreover, on the light-cone 
the correlator is found to decay as $R^{(0)}(r=2T_m) \sim r^{-\frac{\gamma^2K_0}{8}+3\frac{K_{tr}}{4}}$.
When $V_{sg}\neq 0$, using RG improved perturbation theory, we will show that this distinct behavior 
of $R$ outside, on and inside the light-cone survives with exponents that differ from the ones above.
It is interesting to
contrast this behavior with that of a correlator in a quench from an initially gapped state~\cite{Calabrese06,Calabrese11}.
For the latter the horizon effect is more
pronounced as $R$ is exponentially suppressed in position outside the light-cone. In contrast for our case, where the quench is
from an initially gapless state, the change in the behavior of $R$ from outside the light-cone to inside is
less dramatic as the correlators both outside the light-cone and inside decay as power-laws in position. 

Before we give results for the
correlator, we first discuss the $\beta$-function obtained from gradually lowering the cut-off from the
bare value of $\Lambda_0$ to $\Lambda_0/l$. The actual derivation is presented in Section~\ref{beta}.  
Integrating out the fast modes generates corrections not only
to the Luttinger liquid parameters of the Hamiltonian after the quench ($H_{f0}$), 
but also generates new terms whose physical meaning is inelastic
scattering. 
Moreover the $\beta$-function depends in general on the time after the quench because  
the strength of the corrections arising due to the commensurate potential 
together with the effect of the $K_0 \rightarrow K$ quench depend on this time. Expressing
time in units of the cut-off, the $\beta$-function is found to be~\cite{Mitra12b}
\begin{eqnarray}
\frac{dg}{d\ln{l}} = g\left[2-\left(K_{neq} + \frac{K_{tr}}{1+4 T_m^2}\right)\right]
\label{rg1}\\
\frac{dK^{-1}}{d\ln{l}} = \frac{\pi g^2 \gamma^2}{4}I_K(T_m)\label{rg2}\\
\frac{d\eta}{d\ln{l}} = \eta + \frac{\pi g^2\gamma^2 K}{2}I_{\eta}(T_m)\label{rg4}\\
\frac{d(\eta T_{eff})}{d\ln{l}} = 2 \eta T_{eff} + \frac{\pi g^2\gamma^2 K}{4}I_{T_{eff}}(T_m)
\label{rg5}\\
\frac{dT_m}{d\ln{l}} = -T_m \label{rg6}
\end{eqnarray}
Above $T_{m}$ is the rescaled time after the quench where the rescaling with the changing cut-off 
in Eq.~(\ref{rg6}) is similar to the
rescaling of position or frequency, the latter applicable for a system in steady-state.
Moreover Eq.~(\ref{rg6}) implies that $T_m$ is related to the bare 
physical time $T_{m0}$ as $T_{m}=\frac{T_{m0}}{l}$. 
This relation suggests that the time after the quench qualitatively 
acts as an inverse UV cutoff so that the larger is the time after the quench, the more important is the role
of the long wavelength modes on the dynamics~\cite{Mathey09}. 
\begin{figure}
\centering
\includegraphics[width=8cm]{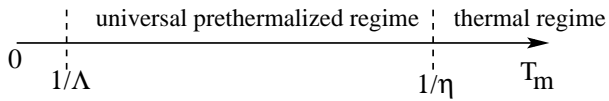}
\caption{The dynamics after a quench is characterized by three different regimes, a short time regime 
that depends on microscopic parameters ($T_m\ll 1/\Lambda$),
an intermediate time regime ($1/\Lambda \ll T_m\ll 1/\eta$) where inelastic effects are weak and the correlator shows universal 
scaling behavior, and a long time thermal regime ($T_m\gg 1/\eta$) where inelastic scattering effects are strong and cause the
system to thermalize.}
\label{fig3}
\end{figure}

Eq.~(\ref{rg2}) represents the usual corrections to the
Luttinger parameter that arise even in the equilibrium theory.
Eq.~(\ref{rg1}) shows that there is a crossover from a short time regime where the scaling dimension of $V_{sg}$ is 
$K_{neq}+K_{tr}-2=\frac{\gamma^2K_0}{4}-2$ and therefore
depends on the initial Luttinger parameter $K_0$, to a long time steady-state regime where the periodic potential 
$V_{sg}$ has a nonequilbrium
scaling dimension $K_{neq}-2$. In the ground state of $H_f$, the scaling dimension of $V_{sg}$ is  $K_{eq}-2$, and since 
$K_{eq}<K_{neq}$, the periodic potential at long times is always more irrelevant 
for the nonequilibrium problem. Thus the location of the critical point at steady-state ($T_m=\infty$) shifts from $K_{eq}=2$ to
$K_{neq}=2$. Physically this is because the interaction quench $K_0 \rightarrow K$ gives rise to a highly excited
state of bosons which cannot be localized by the periodic potential as easily as when they are 
in the zero-temperature ground state. A consequence of the change in the scaling dimension with time from $K_{neq}+K_{tr}-2$ to
$K_{neq}-2$ is a qualitative change in the two-point correlation function from outside the light-cone
to inside the light-cone, an effect which will be discussed in detail later.

Equations~(\ref{rg4}),~(\ref{rg5}) show that the notion of relevance or irrelevance should not be taken 
literally out of equilibrium as
even when the cosine potential is irrelevant, it can cause the generation of new terms after a quantum quench. These terms are 
an inelastic scattering rate $\eta$ which can be identified by the generation of quadratic corrections
to the Luttinger liquid theory of the form $\eta(T_m)\phi_q \partial_{T_m} \phi_{cl}(T_m)$. The second new term 
is a noise for the long wavelength modes of strength $\eta(T_m) T_{eff}(T_m)$ which corresponds  
to the generation of 
corrections to the Luttinger liquid theory of the form $\eta(T_m) T_{eff}(T_m) \phi_q^2$.  Note that for $T_m=0$, all the corrections 
$I_{K,\eta,T_{eff}}$ vanish as the effect of $V_{sg}$ vanishes, while for $T_m\gg 1$, $I_{K,\eta,T_{eff}}$
take steady-state values implying that   
the dissipation and noise reach steady-state values. Thus at long times, the low energy effective theory is a classical theory 
characterized by an effective-temperature $T_{eff}$, and a classical fluctuation-dissipation theorem is obeyed with a
dissipation strength of~\cite{Mitra11} $\eta$. 

An interaction quench in a Luttinger liquid ($V_{sg}=0$) generates a highly nonequilibrium 
occupation of the bosonic modes which does
not relax. However in the presence of $V_{sg}$, the occupation probability of these bosonic modes is no longer conserved, and a nonzero
$\eta$ represents the rate at which the occupation probability of the long-wavelength modes relax. Here by $\eta$
we imply the dissipation strength at long times ($T_m\gg 1$).  
Thus $1/\eta$ is a natural time-scale associated with the quench, which is the time after which inelastic scattering events 
become strong and cause a significant deviation of the bosonic occupation probabilities. Thus the 
$\beta$ function implies that the dynamics
after a quench  has three regimes shown in Fig.~\ref{fig3}. A short time regime $T_m \ll 1$ where
the dynamics depends on microscopic details and can be
easily treated within perturbation theory. The second is an intermediate time prethermalized 
regime where the time is long as compared to microscopic time-scales,
while short as compared to the dissipation-rate $1\ll T_m\ll 1/\eta$. In this regime inelastic effects are weak. 
Using RG improved perturbation theory, we will show that 
the two-point correlation function $R$ shows 
universal behavior in this intermediate time regime, where the precise universal
behavior also depends on the magnitude of the spatial separation relative to the time after the quench (horizon effect).
Finally there is a third regime which we label the thermal regime $T_m\gg 1/\eta$ 
where inelastic effects are strong and lead to eventual thermalization. 
Since for small quenches ($|K_0- K|\ll 1$), $\eta \sim g^2 (K_0-K)^4 \ll 1$~\cite{Mitra12a}, the intermediate
time prethermalized regime can be quite large. In this paper we will give results for the correlation function
in this intermediate time regime where universal dynamics characterized by a CS-like differential equation will emerge.
The dynamics in the thermal regime is also interesting to explore, and will be discussed elsewhere. 

Note that in the nonequilibrium problem, the term irrelevant simply implies that the strength of the perturbation
decreases under RG transformations so that perturbation theory in $g$ is valid. In addition, 
the meaning of the leading irrelevant operator in the nonequilibrium problem is the same as in equilibrium in that
it is the coupling constant that decreases under RG the slowest. For example, under RG (and at long times), the
coupling strength $g$ for the potential $\cos(\gamma\phi)$, according to Eq.~(\ref{rg1}), decreases as $l^{2-K_{neq}}$. This is a slower
decrease than that for the perturbation of the form $\cos(2 \gamma \phi)$ which decreases as $l^{2-4 K_{neq}}$. Under 
RG transformation, the coupling constant for the band curvature $\left(\partial_x \phi\right)^3$ 
will also decrease faster than that of $V_{sg}$ for the values of $K_{neq}$
that we are concerned with in this paper. 
Thus while these other irrelevant terms will also give rise to additional
inelastic scattering, these will only be small corrections to the inelastic scattering rate already produced by
$V_{sg}$.

Eq.~(\ref{rg1}) shows that in the prethermalized regime, 
there is a crossover from an intermediate time dynamics where the physics
is determined by the initial wave-function (and hence the initial Luttinger parameter $K_0$) and a long time
time dynamics determined by $K_{neq}$. This can result in a situation where perturbation theory 
in $g$ is violated at intermediate times when $\frac{\gamma^2K_0}{4}<2$ {\sl i.e.}, when $g$ is a relevant perturbation in the
initial state. In Section~\ref{pertR} and~\ref{latticeint} 
we show how this happens and what it implies. 
Throughout this paper we will present results close to the critical point defined by 
\begin{eqnarray}
K_{neq}= 2+\delta, \,\,\forall \,\, 0<\delta \ll 1 
\end{eqnarray}
where $V_{sg}$ is marginal and can give rise to logarithmic corrections to scaling. 

We now outline how the CS-like differential equation is derived, this
is a summary of the more detailed calculations presented in Section~\ref{CSder}.
We derive the CS-like differential equation for $R$ by integrating out fast modes gradually and in
the process lowering the cut-off from $\Lambda \rightarrow \Lambda/l= \Lambda-d\Lambda$. This leads to a relation of the
form
\begin{eqnarray}
R = R_<\left[1-\frac{d\Lambda}{\Lambda}\left(\ldots\right)\right]
\end{eqnarray}
where $R_<$ is the correlator for the slow modes while $R$ is the correlator for all the modes.
To get an idea for what to expect, let us carry out this exercise for the quadratic theory after the quench ($g=0$). 
Here for time $T_m$ after the quench the following relation between the correlator for the full and the slow modes
emerges
\begin{eqnarray}
R^{(0)}(r,T_m) = R^{(0)}_<(r,T_m)\left[1-\frac{d\Lambda}{\Lambda}\gamma_{an,0}(r,T_m)\right]
\label{R0}
\end{eqnarray}
The above expression 
implies the following differential equation 
$\left[\frac{\partial}{\partial\ln{l}} - \gamma_{an,0}(r,T_m)\right]R^{(0)}\left(\frac{r\Lambda_0}{l},\frac{T_{m0}\Lambda_0}{l}\right) = 0$ 
where (in units of $\Lambda$)
\begin{eqnarray}
\gamma_{an,0}(r,T_m) &&= \frac{1}{2}\left[K_{neq}\frac{r^2}
{1+r^2} + \frac{K_{tr}}{1+(2T_m)^2}
\right.\nonumber \\
&&\left. -\frac{K_{tr}}{2}\biggl\{\frac{1}{1+(2T_m+r)^2}
+\frac{1}{1+(2T_m-r)^2}\biggr\} 
\right]\nonumber \\\label{gan0}
\end{eqnarray}
Eq.~(\ref{gan0}) shows that 
there are three scaling limits where $\gamma_{an,0}$ becomes
independent of $r,T_m$. One is within the light-cone $2T_m \gg r\gg 1$ where $\gamma_{an,0}=K_{neq}/2$. The
second is outside the light-cone $r \gg 2T_m \gg 1$ where also $\gamma_{an,0}=K_{neq}/2$, while 
the third is on the light-cone $r=2T_m,r\gg 1$ where $\gamma_{an,0}= \left(K_{neq}-K_{tr}/2\right)/2$.
We now discuss the correction to $R$ to next order in $g$ where logarithmic corrections
arise in the vicinity of the critical point. 

At next order in the cosine potential (at ${\cal O}(g)$), in terms of slow and fast fields, we find
\begin{eqnarray}
&&R^{(1)}(r,T_m)= R_<^{(1)}(r,T_m)
\left[1 + 2\frac{d\Lambda}{\Lambda} - 2 K_{neq} \frac{d\Lambda}{\Lambda} \right. \nonumber \\
&&\left. + \gamma_{an,0}(r,T_m) \frac{d\Lambda}{\Lambda}\right] 
+ 2 g \pi I_C(r,T_m) \frac{d\Lambda}{\Lambda}R_<^0(r,t)\label{R1}
\end{eqnarray}
where $I_C$ is discussed below.
To quadratic order, the correction $R^{(2)}$ leads to the $\beta$ function which has been discussed above.
Eqs.~(\ref{R0}),~(\ref{R1}) and the $\beta$ function imply the following CS like differential equation for $R$,
\begin{eqnarray}
&&\left[\frac{\partial}{\partial \ln{l}} + \beta(g_i) \frac{\partial }{\partial g_i}- \gamma_{an,0} + 2\pi g I_C\right]\nonumber \\
&&\times R\left[\frac{r\Lambda_0}{l}, 
\frac{\Lambda_0T_{m0}}{l},g_i(l)\right]=0
\label{cz}
\end{eqnarray}
above $T_{m0}$ is the time after the quench, while $r$ is the spatial separation between the local operators.  
$g_i = g,\delta,\eta,T_{eff}$, are coupling constants while $\gamma_{an,0}-2\pi g I_C$ is the anomalous scaling dimension of the
correlator. For macroscopic lengths and times
where $\frac{r\Lambda_0}{l}\gg 1, \frac{T_{m0}\Lambda_0}{l}\gg 1$, $I_C,\gamma_{an,0}$ are constants independent of
position and time. 

We make a simplifying assumption of being in the prethermalized regime $1\ll T_{m0}\ll 1/\eta$. Here the new coupling constants
related to dissipation and noise may be neglected and the $\beta$ function becomes much simpler.
On integrating Eq.~(\ref{cz})
upto $l^* = \Lambda_0 {\rm min}(r,T_{m0})$, one may relate 
the correlator at long times and distances to the correlator at short times and distances 
and a renormalized coupling $g_i(l^*)$, 
where since $g(l^*)\ll 1$, the latter may be evaluated readily within perturbation theory. 
The anomalous scaling dimension $\gamma_{an,0}-2\pi g I_C$
takes different values in the three regimes shown in Fig.~\ref{fig2}, and is responsible for the distinctly different scaling
behavior outside, on and inside the light-cone. We now present results for the correlation function for two cases, one
is for the pure-lattice quench ($K_0=K,g\neq 0$), and the second is a simultaneous lattice and interaction quench ($K_0\neq K, g\neq 0$).

{\bf Pure lattice quench:} This corresponds to $K_0=K$ or $K_{neq}=K_{eq}$, but a periodic potential of strength $g$
switched on suddenly at $T_m=0$.
We are interested in the physics in the vicinity of the critical point where  $K_{eq}=K_{neq}=2+\delta\,\, \forall \,\,0< \delta \ll 1$. Here we find,
\begin{eqnarray}
&&I_C(r,T_m)=\frac{r^2+1}{4 + r^2}  \nonumber \\
&&-\left(\frac{1+r^2}{1 + T_m^2}\right)
\left[\frac{r^2-4T_m^2 + 4 -8T_m^2}{\biggl\{r^2 - 4 T_m^2 + 4\biggr\}^2 + 64 T_m^2}\right]\label{IC1}
\end{eqnarray}
Note that $I_C(T_m=0)=0$ as the lattice has not had time to affect the correlator.
Eq.~(\ref{IC1}) shows the appearance of scaling in three cases, one is outside the light-cone where,
\begin{eqnarray}
I_C(r,T_m\gg 1, 2T_m \ll r) &&=1 + {\cal O}\left(\frac{1}{r^2},\frac{1}{T_m^2}\right)
\end{eqnarray}
the second is within the light-cone,
\begin{eqnarray}
I_C(r,T_m\gg 1, 2T_m \gg r) &&=1 + {\cal O}\left(\frac{1}{r^2},\frac{r^2}{T_m^4}\right)
\end{eqnarray}
and the third is on the light-cone,
\begin{eqnarray}
I_C(r,T_m\gg 1, 2T_m  = r) &&=  \frac{3}{2} + {\cal O}\left(\frac{1}{r^2}\right) \label{IClc}
\end{eqnarray}
Eq.~(\ref{gan0}),~(\ref{IC1}) also show that scaling is valid until (restoring units) $l\sim{\rm min}\left[\Lambda_0 r, \Lambda_0 T_{m}\right]$.
In the scaling limit, $I_K$ in Eq.~(\ref{rg2}) is 
$I_K(T_m\gg 1) = \pi \left[1-\frac{7}{8T_m^2} + \ldots\right]$. Thus
the solution of the CS equation~(\ref{cz}) in the three scaling limits
where $I_C,\gamma_{an,0},I_K$ are constants in time and position is,
\begin{eqnarray}
&&R\left(\Lambda_0 r,\Lambda_0T_{m0},g_0\right) = e^{-\int_{g_0}^{g(l)} dg^{\prime}\frac{\gamma_{an}(g^{\prime})}{\beta(g^{\prime})}}\nonumber \\
&&\times R\left(\frac{r\Lambda_0}{l}, \frac{T_{m0}\Lambda_0}{l},g(l)\right)\label{czsol}
\end{eqnarray}
where 
$\gamma_{an}=1+\frac{\delta}{2}-2\pi g I_C $
and the $\beta$ function is $\frac{dg}{d\ln{l}}=-g\delta,\frac{d\delta}{d\ln{l}}= -(2\pi g)^2$.
Note that the critical point corresponding to the $S=1/2$ Heisenberg chain corresponds to $\delta = 2\pi g$.
Eq.~(\ref{czsol}) is one of the main results of this paper. It shows the existence of a scaling regime where the
correlator at large times or distances is related to the correlator at shorter scales ($r/l,T_{m0}/l$) and renormalized couplings
$g(l)$, where since $g(l\gg 1)\ll 1$, the latter may be readily evaluated within perturbation theory. 
We first discuss the behavior of the correlator $R$ at the critical point $\delta=2\pi g$, and then discuss its behavior
for slight deviations from this critical point such that $\delta > 2\pi g$. 

The correlation function outside the light-cone is determined by setting $l=\Lambda_0 T_{m0}$ in Eq.~(\ref{czsol}), and using
$R\left(\frac{r}{2T_{m0}}\gg 1,l=\Lambda_0T_{m0},g=0\right)\sim \frac{T_{m0}}{r}$ (see Section~\ref{quadratic}). At the critical
point $\delta=2\pi g $ this gives,
\begin{eqnarray}
R\left(r \gg 2T_{m0}\right)\sim \frac{\sqrt{\ln{T_{m0}}}}{r}\label{czsol2}
\end{eqnarray}
The correlation function within and inside the light-cone is obtained by setting $l=\Lambda_0r$ in Eq.~(\ref{czsol}),
Moreover noting that $R\left(l=r\Lambda_0,\frac{T_{m0}}{r}\gg 1,g=0\right)\sim {\cal O}(1)$ this gives the following
correlator inside the light-cone at the critical point,
\begin{eqnarray}
R(r,2T_{m0} \gg r) \sim \frac{\sqrt{\ln{r}}}{r}\label{czsol3}
\end{eqnarray}
In a similar way, using $R\left(l=r\Lambda_0,\frac{2T_{m0}}{r}= 1,g=0\right)\sim {\cal O}(1)$, 
the correlator on the light-cone is found to be
\begin{eqnarray}
R(r=2T_{m0}) = \frac{\ln{r}}{r}\label{czsol4}
\end{eqnarray}
Equations~(\ref{czsol2}),~(\ref{czsol3}),~(\ref{czsol4}) are the main results for the correlator for a
pure lattice quench at the critical point $\delta=2\pi g$. Thus in the prethermalized regime 
the result Eq.~(\ref{czsol3}) within the light-cone for a pure lattice
quench is the same as in the ground state of the Heisenberg chain~\cite{Singh89,Affleck98,Barzykin99}. In contrast, the quench
leads to qualitatively new scaling behavior for spatial separations outside the light-cone (Eq.~(\ref{czsol2})) and 
on the light-cone (Eq.~(\ref{czsol4})).  
Note that for a pure lattice quench, no dissipative effects are generated 
to ${\cal O}(g^2)$~\cite{Mitra12a}, extending the regime of validity of the prethermalized regime.  

For slight deviations $\delta > 2\pi g$ from the critical point, we obtain the following correlator outside the
light-cone 
\begin{eqnarray}
&&R\left(\Lambda_0r,\Lambda_0T_{m0},g_0, r \gg 2T_{m0}\right)\sim \nonumber \\
&&\left(\frac{1}{r}\right)^{1+\delta/2}\left(T_{m0}\right)^{\frac{\delta}{2}-\frac{\sqrt{\delta^2-(2\pi g)^2}}{2}}
\label{Rolca}
\end{eqnarray}
Thus the correlator outside the light-cone is primarily the one in the initial state with a time-dependent prefactor
which depends on the strength $g$ of the periodic potential. Eventually this time-dependence drops off at very long times,
with the correlator taking the following steady-state value inside the light-cone,
\begin{eqnarray}
R\left(\Lambda_0r,\Lambda_0T_{m0},g_0, r \ll 2T_{m0}\right)\sim \frac{1}{r^{1+\frac{1}{2}\sqrt{\delta^2 - (2\pi g)^2}}}\label{Rilca}
\end{eqnarray}
Unlike the case of the dynamics on the critical point, 
for slight deviations from the critical point $\delta > 2\pi g$, 
the leading asymptote for the correlator on and inside the light-cone is the same. 
The expressions for the correlators in Eqs.~(\ref{Rolca}),~(\ref{Rilca}) 
are valid for $\left(\sqrt{\delta^2-(2\pi g)^2}\right)
\ln{\left({\rm min}\left[T_{m0},r\right]\right)}\gg 1$, whereas the
universal logarithmic corrections discussed before this in Eqs.~(\ref{czsol2}),~(\ref{czsol3}),~(\ref{czsol4}) are 
valid for the opposite case of
$\left(\sqrt{\delta^2-(2\pi g)^2}\right)\ln{\left({\rm min}\left[T_{m0},r\right]\right)}\ll 1$.  

{\bf Simultaneous lattice and interaction quench:} Now we turn to the case where both the Luttinger interaction parameter and
the cosine potential is simultaneously quenched at $T_m=0$. Moreover we are interested in the physics close to the nonequilibrium
critical point defined by $K_{neq}=2 +\delta, \forall\,\, 0<\delta \ll 1$. 
This quench corresponds to a final Hamiltonian whose ground state can be in the gapped phase. However due to the quench,
since $K_{neq}>K_{eq}$, the periodic potential is more irrelevant (though it can give rise to inelastic scattering). Thus 
perturbation theory in the cosine potential may be valid for the nonequilibrium problem, even though it may not hold
for determining the properties of the ground state.

\begin{figure}
\centering
\includegraphics[totalheight=4cm]{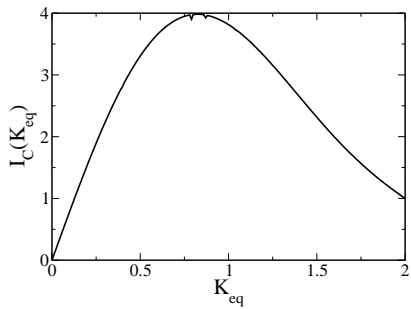}
\caption{Plot of $I_C$ for a simultaneous lattice and interaction quench for points within the light-cone
and near the critical point $K_{neq}=2$ where $K_{eq}=\frac{\gamma^2K_0}{4}\sqrt{\frac{16}{\gamma^2 K_0}-1}$. 
}
\label{figL}
\end{figure}

For this general quench we find that $I_K$ in Eq.~(\ref{rg2}) is 
$I_K(T_m,K_{neq}=2) $=$\pi\left[c_1
-c_2 \sin\left(\pi (K_{eq}-K_{neq})\right)\ln\left({\rm min}\left[T_m,1/\eta\right]\right)\right]$ 
where $c_{1,2}$ are ${\cal O}(1)$ and depend on the initial Luttinger parameter $K_0$. When $K_0$=$K$, $c_1$=$1$. 
As before we are interested in the  prethermalized regime where $1\ll T_m < 1/\eta$. 
We will also assume a small quench where $|K_{eq}-K_{neq}|\ln{T_m} \ll 1$.
In this case, $I_K\simeq \pi c_1$ is a constant in time. 
The $\beta$ function in the vicinity of $K_{neq}=2+\delta$ becomes
$\frac{dg}{d\ln{l}} = -g \delta,
\frac{d\delta}{d\ln{l}} = -g^2B^2$, where $B=\pi\sqrt{c_1} \frac{\gamma^2K_0}{4}\left(\frac{16}{\gamma^2K_0}-1\right)^{3/4}$.
In addition,  $I_C$ in Eq.~(\ref{cz}) becomes a universal function of the initial Luttinger parameter $K_0$ and is plotted  in
Fig.~\ref{figL}.

Using Eq.~(\ref{czsol2}) the correlation function inside the light-cone is found to be 
\begin{eqnarray}
&&R\left(\Lambda_0 r, r \ll 2T_{m0}\ll 1/\eta,g_0\right)\simeq \frac{1}{\left(\Lambda_0 r\right)^{1+A/2}}\nonumber \\
&&\frac{1}{\sqrt{1-(r\Lambda_0)^{-2 A}}}
\left[\frac{1-(\Lambda_0 r)^{-A}}{1+(\Lambda_0r)^{-A}}\right]^{\frac{2\pi I_C}{B}}\label{Rgen}
\end{eqnarray}
where $A=\sqrt{\delta^2-g^2 B^2}$.
Thus for an interaction and 
lattice quench, and for $A \ln{r} \gg 1$, the correlators decay as a power-law with 
exponent ($1+A/2$). This exponent is not the same as in an interaction quench in a quadratic Luttinger liquid 
theory  which would have been $1+\delta/2$. Thus the lattice, even though irrelevant, modifies the decay exponent.
In the vicinity of the nonequilibrium critical point $A\rightarrow 0$, 
logarithmic corrections are obtained for $ A \ll A\ln{r} \ll 1$, 
where 
\begin{eqnarray}
R(2 T_{m0}\gg r)\sim \frac{1}{r}\left(\ln{r}\right)^{\frac{2\pi I_C}{B}-\frac{1}{2}} \label{log3}
\end{eqnarray}
This steady-state behavior in the vicinity of the nonequilibrium critical point is significantly different from that 
near the equilibrium critical point which is~\cite{Singh89,Affleck98,Barzykin99} $\sqrt{\ln{r}}/r$.

Let us briefly discuss scaling in the other two regimes, one is on the light-cone and the other is outside
the light-cone. 
For the former we find
\begin{eqnarray}
I_C(r=2T_{m0}) \sim r^{K_{tr}/2}
\end{eqnarray}
where $K_{tr}=\frac{\gamma^2K_0}{4}-2$. Thus scaling is recovered only if $K_{tr}\leq 0$ where $I_C \rightarrow 0$ or to 
a constant at large 
distances. Whereas for $K_{tr}>0$ scaling is lost as $I_C$ grows with distance, only to be cut-off at $r \sim 1/\eta$ where
we expect the correlators to begin decaying in a thermal manner. 

In contrast to the above, outside the light-cone we find,
\begin{eqnarray}
I_C(r\gg 2T_{m0})\sim \left(\frac{2T_{m0}}{r}\right)^{K_{tr}}
\end{eqnarray}
Here when $K_{tr}\geq 0$, scaling holds outside the light-cone as $I_C$ is either a constant or decays to zero for 
sufficiently large distances. On the other hand when $K_{tr}<0$, $I_C$ grows with distance. This behavior
is opposite to what one finds on the light-cone, and is consistent with the fact that the behavior of the correlator outside
the light-cone is primarily determined by the initial wave-function. Thus if the cosine potential is a relevant perturbation
in the initial state $K_{tr}<0$, then the perturbative corrections are large indicating that perturbation theory may not be
valid at large distances outside the light-cone, even though it may be valid inside the light-cone. This behavior is also
consistent with the crossover in time of the scaling dimension of $V_{sg}$ discussed earlier in this section.

The remaining part of the paper outlines how the above results were obtained. In Section~\ref{quadratic} we 
reintroduce the model and briefly present results for an interaction
quench in the quadratic theory (the Luttinger liquid) in the language of Keldysh Green's functions. 
These results will be used in later sections when we do perturbation theory in the cosine
potential. In Section~\ref{beta}, we do perturbation theory in the cosine potential and derive the $\beta$-function to 
two loop. In Section~\ref{pertR} we present results for the correlation function within perturbation theory to leading order in
the cosine potential. These results set the stage for doing renormalization improved perturbation theory which will be
explicitly carried out in Section~\ref{CSder} where the CS equation for the correlation function is
derived. Results of solution of the CS equation is presented in Section~\ref{lattice} for the case where only the
lattice potential is quenched, while results for the correlation function for a simultaneous lattice and interaction
quench are presented in Section~\ref{latticeint}. Finally in Section~\ref{summary} we summarize our results
and discuss open questions.

\section{Model and Green's functions for the quadratic theory} \label{quadratic}

In order to study quench dynamics of the spin-chain, we employ a bosonization prescription where,
\begin{eqnarray}
&&\phi(x) =  
-(N_{R}+N_{L})\frac{\pi x}{L}\nonumber \\
&&-\frac{i\pi}{L}\sum_{p\neq0}\left(\frac{L|p|}{2\pi}\right)^{1/2}
\frac{1}{p}
e^{-\alpha|p|/2-ipx}\left(b_{p}^{\dagger} + b_{-p}\right), \label{z1}\\
&&\theta(x) = 
(N_{R}-N_{L})\frac{\pi x}{L} \nonumber \\+ 
&&\frac{i\pi}{L}\sum_{p\neq0}
\left(\frac{L|p|}{2\pi}\right)^{1/2}
\frac{1}{|p|}e^{-\alpha|p|/2-ipx}\left(b_{p}^{\dagger} - b_{-p}\right). \label{z2}
\end{eqnarray}

We choose the initial Hamiltonian for $t\leq 0$ to be a Luttinger liquid,
\begin{eqnarray}
&&H_i = \frac{u_0}{2\pi}\int dx
\left[K_0\{\pi \Pi(x)\}^2 + \frac{1}{K_0}\{\partial_x \phi(x)\}^2
\right] \nonumber \\
&&= \sum_{p\neq 0} u_0 |p| \eta_p^{\dagger} \eta_p
\label{Hidef}
\end{eqnarray}
while the time evolution from $t >0$ is due to the quantum sine-Gordon model,
\begin{eqnarray}
&& H_f = H_{f0} + V_{sg}\\
&&H_{f0} = \frac{u}{2\pi}\int dx
\left[K\{\pi \Pi(x)\}^2 + \frac{1}{K}\{\partial_x \phi(x)\}^2
\right] \nonumber \\
&&= \sum_{p\neq 0}u|p|\gamma_p^{\dagger}\gamma_p
\label{Hfdef}\\
&& V_{sg} = -\frac{gu}{\alpha^2} \int dx \cos(\gamma \phi)
\end{eqnarray}
We will make the assumption that the quench connects the same 
zero-mode sectors of the initial and final Hamiltonian. In this case, the zero modes 
corresponding to the first terms in Eq.~(\ref{z1}),~(\ref{z2}) 
will not play a role in the dynamics. 

We study a quench that preserves Galilean invariance {\sl i.e.},
$u = v_F/K, u_0=v_F/K_0$. At the microscopic
level, this corresponds to a quench in the Luttinger model
where the $g_2$ and $g_4$ interactions equal each other for both the initial and final Hamiltonians
($g_{2i}=g_{4i}, g_{2f}=g_{4f}$)~\cite{Giamarchibook}.
While this simplifies the algebra, relaxing this requirement is straightforward, and does not change 
the results in a qualitative way.
The three bosonic operators $b,\eta,\gamma$ are related by a linear Bogoliubov transformation,
\begin{eqnarray}
\begin{pmatrix} b_p \\ b_{-p}^{\dagger}\end{pmatrix}
= \begin{pmatrix} \cosh\beta & -\sinh\beta \\-\sinh\beta &\cosh\beta
\end{pmatrix}
\begin{pmatrix} \gamma_p \\ \gamma_{-p}^{\dagger}\end{pmatrix}\\
\begin{pmatrix} b_p \\ b_{-p}^{\dagger}\end{pmatrix}
= \begin{pmatrix} \cosh\beta_0 & -\sinh\beta_0 \\-\sinh\beta_0 &\cosh\beta_0
\end{pmatrix}
\begin{pmatrix} \eta_p \\ \eta_{-p}^{\dagger}\end{pmatrix}
\end{eqnarray}
where $e^{-2\beta_0}=K_0, e^{-2\beta} = K$.

Let us define the functions
\begin{eqnarray}
f(pt) &&= \cos(u\mid p\mid t)\cosh\beta_0 \nonumber\\
&&-i \sin(u\mid p\mid t)\cosh(2\beta-\beta_0)\\
g(pt) &&= \cos(u\mid p\mid t)\sinh\beta_0 \nonumber \\
&&+ i \sin(u\mid p\mid t)\sinh(2\beta-\beta_0)
\end{eqnarray}
which determine the time-evolution after the quench ($t > 0$) for the quadratic theory ($g=0$),
\begin{eqnarray}
b_p^{\dagger}(t) + b_{-p}(t) &&=\left(f^*(pt)-g(pt)\right)\eta_p^{\dagger}(0) \nonumber\\
&&+
\left(f(pt)-g^*(pt)\right)\eta_{-p}(0)\\
b_p^{\dagger}(t) - b_{-p}(t) &&=\left(f^*(pt)+g(pt)\right)\eta^{\dagger}_p(0) \nonumber \\
&&-\left(f(pt)+g^*(pt)\right)\eta_{-p}(0)
\end{eqnarray}

Since the system is out of equilibrium, it is convenient to study the problem using the Keldysh formalism. 
The Keldysh action is,
\begin{eqnarray}
&&Z_K =Tr\left[\rho(t)\right]= Tr\left[e^{-iH_f t}|\psi_i\rangle\langle\psi_i|e^{i H_f t}\right]\\
&&= \int {\cal D}\left[\phi_{cl},\phi_q\right] e^{i \left(S_0 + S_{sg}\right)}
\end{eqnarray}
where $S_0$ is the quadratic part which describes the physics in the absence of the periodic
potential which corresponds to an interaction quench in a Luttinger liquid. 
In particular at a time $t$ after the quench (note that the fields $\phi$ are real),
\begin{eqnarray}
&&S_0 = \frac{1}{2}\int_{-\infty}^{\infty} dx_1 \int_{-\infty}^{\infty} dx_2\int_0^{t} dt_1\int_0^t dt_2
\begin{pmatrix} \phi_{cl}(1) & \phi_q(1)\end{pmatrix}\nonumber \\
&&\begin{pmatrix} 0&&G_A^{-1}(1,2)\\
G_R^{-1}(1,2) && -\left[G_R^{-1} G_K G^{-1}_A\right](1,2)
\end{pmatrix}\begin{pmatrix}
\phi_{cl}(2)\\
\phi_q(2)
\end{pmatrix}\label{S0}
\end{eqnarray}
where $1=(x_1,t_1), 2= (x_2,t_2)$ and
\begin{eqnarray}
\phi_{cl,q}=\frac{\phi_-\pm \phi_+}{\sqrt{2}}
\end{eqnarray}
where $G_{R,A,K}$ are the retarded, advanced and Keldysh Green's functions, with
\begin{eqnarray}
&&\left[G_{R,A}(1,2)\right]^{-1} = -\delta(x_1-x_2)\delta(t_1-t_2)\nonumber \\
&&\times \frac{1}{\pi K u}
\left[\partial^2_{t_1\pm i\delta}-u^2\partial_{x_1}^2\right]
\end{eqnarray}
and,
\begin{eqnarray}
-i\langle \begin{pmatrix}
\phi_{cl}(1)\\\phi_q(1) \end{pmatrix}
\begin{pmatrix} \phi_{cl}(2) & \phi_q(2)\end{pmatrix} \rangle
= \begin{pmatrix} G_K(1,2) & G_R(1,2) \\ G_A(1,2) & 0\end{pmatrix}\nonumber \\
\end{eqnarray}
Whereas,
\begin{eqnarray}
&&S_{sg} = \nonumber \\
&&\frac{gu}{\alpha^2}\int_{-\infty}^{\infty}dx \int_0^t dt_1\left[\cos\{\gamma\phi_-(1)\}
-\cos\{\gamma\phi_{+}(1)\}\right]
\end{eqnarray}

For the quadratic theory after the quench,
\begin{eqnarray}
&&G_R(x_1t_1,x_2t_2) = -i\theta(t_1-t_2)\langle \left[\phi(x_1t_1),\phi(x_2t_2)\right] \rangle\\
&&=-\frac{K}{2}\theta(t_1-t_2)
\sum_{\epsilon=\pm}\tan^{-1}\left(\frac{u(t_1-t_2)+\epsilon(x_1-x_2)}{\alpha}\right)
\nonumber \\
&&G_A(x_1t_1,x_2t_2) =  i\theta(t_1-t_2)\langle \left[\phi(x_1t_1),\phi(x_2t_2)\right]\\
&&= \frac{K}{2}\theta(t_2-t_1)
\sum_{\epsilon=\pm}\tan^{-1}\left(\frac{u(t_1-t_2)+\epsilon(x_1-x_2)}{\alpha}\right)\nonumber 
\end{eqnarray}
and, 
\begin{eqnarray}
&&G_K(x_1t_1,x_2t_2) = -i\langle \{\phi(x_1t_1),\phi(x_2t_2)\}\rangle\\
&&= -i\frac{K_0}{4}\left(1+\frac{K^2}{K_0^2}\right)\int_0^{\infty}\frac{dp}{p}e^{-\alpha p}\nonumber\\
&&\times \sum_{\epsilon=\pm}\cos(up(t_1-t_2)
+\epsilon p(x_1-x_2)) \nonumber \\
&&-i\frac{K_0}{4}\left(1-\frac{K^2}{K_0^2}\right)\int_0^{\infty}\frac{dp}{p}e^{-\alpha p}\nonumber \\
&&\times \sum_{\epsilon=\pm}\cos(u p(t_1+t_2)+\epsilon p (x_1-x_2))
\end{eqnarray}
Note that $G_K$ is logarithmically divergent, but in all physical quantities it is always the combination
$G_K(1,2)-\frac{1}{2}G_{K}(1,1)-\frac{1}{2}G_K(2,2)$ that appears, which is finite.

Let us define
\begin{eqnarray}
&&C_{ab,m}(x_1t_1,x_2t_2) = 
\langle e^{i m \gamma \phi_a(x_1t_1)}e^{-i m \gamma \phi_b(x_2t_2)}\rangle =\nonumber \\
&&e^{-\frac{\gamma^2 m^2}{2}\left[\frac{iG_K(11)}{2} + \frac{iG_K(22)}{2}-i G_K(12)+ i a G_A(1,2)+i b G_R(12)\right]}
\end{eqnarray}
where
\begin{eqnarray}
&&-\frac{\gamma^2}{2}\left[\frac{iG_K(1,1)}{2}+ \frac{iG_K(2,2)}{2}-iG_K(1,2)\right]\nonumber \\
&&= -K_{neq}\sum_{\epsilon=\pm}\left[\ln{\frac{\sqrt{\alpha^2 + \{u(t_1-t_2)+\epsilon(x_1-x_2)\}^2}}{\alpha}}
\right]\nonumber\\
&&-K_{tr} \left[
\ln\sqrt{\frac{\alpha^2+\{u(t_1+t_2)+(x_1-x_2)\}^2}{\alpha^2+(2ut_1)^2}}\right.\nonumber \\
&&\left. +\ln\sqrt{\frac{\alpha^2+\{u(t_1+t_2)-(x_1-x_2)\}^2}{\alpha^2+(2ut_2)^2}} \right]\label{exp1}
\end{eqnarray}
with the coefficients $K_{eq},K_{neq},K_{tr}$ defined in Eq.~(\ref{Kexp}).
The above implies that the equal-time correlator is given by (setting $u=1$ and $\Lambda=1/\alpha$)
\begin{eqnarray}
&&C_{m=1}(rt,0t) = \frac{1}{\left(\sqrt{1+ \Lambda^2 r^2}\right)^{2K_{neq}}}\nonumber \\
&&\times \left[
\frac{\sqrt{1+\Lambda^2(2t+r)^2}\sqrt{1+\Lambda^2(2t-r)^2}}{1 + \left(2 \Lambda t\right)^2}\right]^{-K_{tr}}\label{Cfull}
\end{eqnarray}
At equal times $C_{ab,m}$ does not depend on the $ab$ indices since $G_{R,A}(t,t)=0$. Therefore
we have dropped the $ab$ indices.

At positions and times large as compared to the cut-off and 
far from the light-cone $|r\pm 2t|\gg 1$ (with $r,t$ measured in units of $\Lambda$)
\begin{eqnarray}
C_{m=1}(rt,0t) = \left[\frac{1}{r^2}\right]^{\frac{\gamma^2K_0}{4}}\left[\lvert \frac{r^2 - (2t)^2}{r^2(2t)^2}\rvert\right]^{-K_{tr}}
\end{eqnarray}
This agrees with Ref.~\onlinecite{Iucci09} where it was pointed out that outside the light-cone ($2t \ll r$) the correlator decays in position in the same 
way as in the initial state, but with a time-dependent prefactor which goes as $t^{2K_{tr}}$. Moreover within the light-cone
$r \ll 2t$, the correlator reaches a steady-state value where it decays in position with the new nonequilibrium
exponent $K_{neq}$, $C_{m=1}(r\ll 2 t) \sim \frac{1}{r^{2K_{neq}}}$. 
Exactly on the light-cone $C_{m=1}(r=2t)\simeq  \left[\frac{1}{r^2}\right]^{\frac{\gamma^2K_0}{4}}\left[ \frac{1}{r^3}\right]^{-K_{tr}} $.

In this paper we aim to calculate the correlator $R_{ab} =\langle e^{i\gamma\phi_a(1)/2}e^{-i\gamma\phi_b(2)/2}\rangle
= C_{ab,m=1/2}$ at equal times and unequal positions in the presence of the cosine potential.
In particular we will derive a CS-like differential equation that will allow us to relate the correlator at large scales of the 
bare theory, to the correlator at short scales and renormalized couplings, where the latter is well approximated by the correlator of the free theory. 
Three useful results for the free correlator at short-scales 
that will be used later are for points within the light-cone $r \ll 2t, \Lambda r \sim 1$, 
points outside the light-cone $r\gg 2t, \Lambda t \sim 1$ and points on the light-cone $r=2t,\Lambda r\sim 1$.
For these three cases using Eq.~(\ref{Cfull}) we find,
\begin{eqnarray}
C_{m=1/2}(r\ll 2t, \Lambda r = 1)\sim {\cal O}(1)\label{Csd1}\\
C_{m=1/2}(r= 2t, \Lambda r = 1)\sim {\cal O}(1)\label{Csd2}\\
C_{m=1/2}(r \gg 2t, \Lambda t = 1)\sim \left(\frac{t}{r}\right)^{K_{neq}/2}\left(\frac{r}{t}\right)^{-K_{tr}/2}\label{Csd3}
\end{eqnarray}

\section{Derivation of the $\beta$ function from the action} \label{beta}

In this section we discuss how the $\beta$-function is derived from the Keldysh action. 
We split the fields into slow ($\phi^<$) and fast ($\phi^{>}$) components where the
fast components have a large weight at high momentum, and therefore oscillate rapidly
in time~\cite{Mitra11,Mitra12a,Torre12,Mitra12b}
\begin{eqnarray}
\phi_{\pm} = \phi^{<}_{\pm} + \phi^{>}_{\pm}
\end{eqnarray}
We integrate out the fast fields perturbatively in the cosine potential. In doing so to ${\cal O}(g^2)$,
we obtain
\begin{eqnarray}
S = S_0^{<} + \delta S^{<}
\end{eqnarray}
where $S_0^<$ is the quadratic action for the slow fields,
\begin{eqnarray}
&&S_0^{<} = \int_{-\infty}^{\infty} dR \int_0^{ut/\sqrt{2}} d(uT_m)
\frac{1}{2\pi K}\left[\phi_q^<\left(\partial_R^2-\partial_{uT_m}^2\right)\phi_{cl}^<\right. \nonumber \\
&&\left. +\phi_{cl}^<\left(\partial_R^2-\partial_{uT_m}^2\right)\phi_{q}^<
+ \frac{\delta u}{u}\phi_q^<\left(\partial_R^2+\partial_{uT_m}^2\right)\phi_{cl}^< \right. \nonumber \\
&&\left. + \frac{\delta u}{u}\phi_{cl}^<\left(\partial_R^2+\partial_{uT_m}^2\right)\phi_{q}^<\right. \nonumber \\
&&\left. - 2\frac{\eta}{u}\phi_q^<\partial_{uT_m}\phi_{cl}^< +i \frac{4 \eta T_{eff}}{u^2}
\left(\phi_q^<\right)^2\right]\nonumber \\
\end{eqnarray}
Above $\delta u$ represents corrections to the velocity, $\eta$ represents dissipation of the long wavelength modes,
and $\eta T_{eff}$ denotes the strength of the noise on the long-wavelength modes. Initially, 
$\delta u= T_{eff}=\eta=0$, but in the first step of the RG, as we show below, the cosine
potential generates corrections (contained in $\delta S^<$) that not only 
renormalize the interaction parameter $K$ and the velocity $u$, but also 
generates the dissipative term ($\phi_q\partial_{T_m}\phi_{cl}$) and noise term ($\phi_q^2$).

The corrections arising from integrating the fast fields are,
\begin{eqnarray}
&&\delta S^{<} = \frac{g\Lambda^2}{u}\int_{-\infty}^{\infty}dx \int_0^t dt_1
\left[\cos\gamma\phi_-^<(1) - \cos\gamma\phi_+^<(1)\right]\nonumber \\
&&\times e^{-\frac{\gamma^2}{4}\langle\left(\phi_{cl}^>(1)\right)^2\rangle}\\
&&+\frac{ig^2\Lambda^4}{2u^2}\int_{-\infty}^{\infty}dx_1 \int_0^t dt_1\int_{-\infty}^{\infty}dx_2 \int_0^t dt_2
\theta(t_1-t_2)\nonumber \\
&&\times:\cos\left(\gamma\phi_-^<(1) -\gamma\phi_-^{<}(2)\right):
e^{-\frac{\gamma^2}{2}\langle(\phi_-(1)-\phi_-(2))^2\rangle}\nonumber \\
&&\times \left[1-e^{-\gamma^2 \langle \phi_-^>(1)\phi_-^>(2)\rangle}\right]\\
&&+ \frac{ig^2\Lambda^4}{2u^2}\int_{-\infty}^{\infty}dx_1 \int_0^t dt_1\int_{-\infty}^{\infty}dx_2 \int_0^t dt_2\nonumber \\
&&\times \theta(t_2-t_1)
:\cos\left(\gamma\phi_+^<(1) -\gamma\phi_+^{<}(2)\right):
e^{-\frac{\gamma^2}{2}\langle(\phi_+(1)-\phi_+(2))^2\rangle}\nonumber \\
&&\times \left[1-e^{-\gamma^2 \langle \phi_+^>(1)\phi_+^>(2)\rangle}\right]\\
&&-\frac{ig^2\Lambda^4}{2u^2}\int_{-\infty}^{\infty}dx_1 \int_0^t dt_1\int_{-\infty}^{\infty}dx_2 \int_0^t dt_2\nonumber\\
&&\{\theta(t_1-t_2)+ \theta(t_2-t_1)\}\nonumber \\
&&\times :\cos\left(\gamma\phi_+^<(1) -\gamma\phi_-^{<}(2)\right):
e^{-\frac{\gamma^2}{2}\langle(\phi_+(1)-\phi_-(2))^2\rangle}\nonumber \\
&&\times \left[1-e^{-\gamma^2 \langle \phi_+^>(1)\phi_-^>(2)\rangle}\right]
\end{eqnarray}
Above we have used that $\cos(a) = :\cos(a): e^{-\langle a^2\rangle/2}$ where the operators inside the symbol $::$ 
are normal -ordered. Moreover, all expectation values $\langle \ldots \rangle$ are with respect to the initial state,
and therefore depend on the quench.
Since the correlators for the full fields is related to the correlator for the
slow and fast fields as follows $G = G^< + G^>$, 
the correlator for the fast fields may be related in a simple way to derivatives of the full correlators~\cite{Nozieres87},
\begin{eqnarray}
G^> = d\Lambda \frac{d G}{d\Lambda}
\end{eqnarray}

Explicit expressions for the fast correlators at equal time are
\begin{eqnarray}
&&\langle\left[\phi_{cl}^>(t)\right]^2 \rangle = \frac{d\Lambda}{\Lambda}
\left[\frac{K_0}{2}\left(1+\frac{K^2}{K_0^2}\right)\right. \nonumber \\
&&\left. +\frac{K_0}{2}\left(1-\frac{K^2}{K_0^2}\right)
\biggl\{\frac{1}{1+\left(2t\Lambda\right)^2}\biggr\}\right]\\
&&\xrightarrow{t\ll 1/\Lambda } \frac{d\Lambda}{\Lambda}K_0\\
&&\xrightarrow{t\gg 1/\Lambda} \frac{d\Lambda}{\Lambda}
\frac{K_0}{2}\left(1+\frac{K^2}{K_0^2}\right)
\end{eqnarray}
The short and long time limits of $\langle \left[\phi_{cl}^>(t)\right]^2\rangle $ reflect the fact that at short times
after the quench, it is the initial wave-function and hence the initial Luttinger parameter $K_0$ that determines
the behavior of the correlators, while at long times, a new nonequilibrium exponent related to $K_{neq}$ determines
the behavior.

For the non-local fast correlators, we have
\begin{eqnarray}
&&\langle \phi_{cl}^>(1) \phi_{cl}^>(2)\rangle = \frac{d\Lambda}{\Lambda}\sum_{\epsilon=\pm}\nonumber \\
&&\left[
\frac{K_0}{4}\left(1+\frac{K^2}{K_0^2}\right)\frac{1}
{1 + \Lambda ^2((t_1-t_2)+\epsilon(x_1-x_2)/u)^2} \right. \nonumber \\
&&\!\!\left. +
\frac{K_0}{4}\left(1-\frac{K^2}{K_0^2}\right)\!\!\frac{1}
{1 + \Lambda^2((t_1+t_2)+\epsilon(x_1-x_2)/u)^2}\right]\\
&&\langle \phi_{cl}^>(1) \phi_{q}^>(2)\rangle = -i\frac{K}{2}\frac{d\Lambda}{\Lambda}\theta(t_1-t_2)
\nonumber \\
&& \sum_{\epsilon=\pm}\left[\frac{\Lambda((t_1-t_2)+\epsilon(x_1-x_2)/u)}
{1 + \Lambda^2((t_1-t_2)+\epsilon(x_1-x_2)/u)^2}\right]\\
&& \langle \phi_{q}^>(1) \phi_{cl}^>(2)\rangle = i\frac{K}{2}\frac{d\Lambda}{\Lambda}\theta(t_2-t_1)
\nonumber \\
&&\sum_{\epsilon=\pm}\left[\frac{\Lambda((t_1-t_2)+\epsilon(x_1-x_2)/u)}
{1 + \Lambda^2((t_1-t_2)+\epsilon(x_1-x_2)/u)^2}\right]
\end{eqnarray}

In the next step we define new variables corresponding to center of mass ($R,T_m$) and relative coordinates
($r,\tau$),
\begin{eqnarray}
R = \frac{x_1+x_2}{2}\,\,\, , T_m= \frac{t_1+t_2}{2}\\
r = x_1-x_2 \,\,\, , \tau = t_1-t_2
\end{eqnarray}
Thus
\begin{eqnarray}
\int_{-\infty}^{\infty}dx_1 \int_{-\infty}^{\infty}dx_2=\int_{-\infty}^{\infty}dR
\int_{-\infty}^{\infty}dr\\
\int_{0}^{t}dt_1 \int_{0}^{t}dt_2= \int_0^{t/\sqrt{2}} dT_m \int_{-2T_m}^{2T_m} d\tau
\end{eqnarray}
Since quantities have a slower variation with respect to
the center of mass coordinates as compared to the 
relative coordinates, we perform a gradient expansion in $R,T_m$ and obtain
\begin{eqnarray}
\delta S^{<} = \delta S_g^{<}+\delta S_0^{<} + \delta S_{T_{eff}}^{<} + \delta S_{\eta}^{<} 
\end{eqnarray}
where
\begin{eqnarray}
&&\delta S^{<}_g = \frac{gu}{\alpha^2}\int_{-\infty}^{\infty}dx_1 \int_0^t dt_1
\left[\cos\gamma\phi_-^<(1) - \cos\gamma\phi_+^<(1)\right]\nonumber\\
&&\times e^{-\frac{\gamma^2}{4}\langle\left[\phi_{cl}^>(1)\right]^2\rangle}
\end{eqnarray}
and
\begin{eqnarray}
&&\delta S_{0}^{<} = \frac{g^2\Lambda^4\gamma^2}{2u^2}\frac{d\Lambda}{\Lambda}
\int_{-\infty}^{\infty} dR\int_{-\infty}^{\infty}dr \int_0^{t/\sqrt{2}} dT_m \int_{-2T_m}^{2T_m} d\tau \nonumber\\
&&\theta(\tau)\left[\left(r\partial_R \phi_{cl}^<\right) \left(r\partial_R \phi_{q}^<\right)
+ \left(\tau\partial_{T_m} \phi_{cl}^<\right) \left(\tau\partial_{T_m} \phi_{q}^<\right)\right]\nonumber \\
&&\times {\rm Im}\left[e^{-\frac{\gamma^2}{2}\langle \left[\phi_+(R+r/2,T_m+\tau/2)-\phi_-(R-r/2,T_m-\tau/2)\right]^2\rangle}\right. \nonumber\\
&&\left. \times F(r,T_m,\tau)\right] 
\end{eqnarray}
while 
\begin{eqnarray}
&& \delta S_{T_{eff}}^{<}= \frac{ig^2\Lambda^4\gamma^2}{2u^2}\frac{d\Lambda}{\Lambda}
\int_{-\infty}^{\infty} dR\int_{-\infty}^{\infty}dr \int_0^{t/\sqrt{2}} dT_m \int_{-2T_m}^{2T_m} d\tau\nonumber \\
&&\left(\phi_q^{<}(R,T_m)\right)^2\nonumber \\
&&{\rm Re}\left[e^{-\frac{\gamma^2}{2}\langle \left[\phi_+(R+r/2,T_m+\tau/2)-\phi_-(R-r/2,T_m-\tau/2)\right]^2\rangle}\right. \nonumber \\
&&\left. \times F(r,T_m,\tau)\right]
\end{eqnarray}
and 
\begin{eqnarray}
&&\delta S_{\eta}^{<} = \frac{g^2\Lambda^4\gamma^2}{2u^2}\frac{d\Lambda}{\Lambda}
\int_{-\infty}^{\infty} dR\int_{-\infty}^{\infty}dr \int_0^{t/\sqrt{2}} dT_m \int_{-2T_m}^{2T_m} d\tau\nonumber \\
&&\phi_q^<(R,T_m)\tau \partial_{T_m}\phi_{cl}^<(R,T_m)\nonumber \\
&&\times {\rm Im}\left[e^{-\frac{\gamma^2}{2}\langle \left[\phi_+(R+r/2,T_m+\tau/2)-\phi_-(R-r/2,T_m-\tau/2)\right]^2\rangle}\right. \nonumber \\
&&\left. \times F(r,T_m,\tau)\right]
\end{eqnarray}
where
\begin{eqnarray}
&&F(r,T_m,\tau) = K_{neq}\left[\frac{1}{1 + \Lambda^2\left(\tau+r/u\right)^2} + 
\frac{1}{1 + \Lambda^2\left({\tau-r/u}\right)^2}\right]\nonumber \\
&&+ K_{tr}\left[\frac{1}{1 + \Lambda^2\left(2T_m+r/u\right)^2} + \frac{1}{1 + 
\Lambda^2\left({2T_m-r/u}\right)^2}\right]\nonumber \\
&&-i K_{eq}\left[\frac{\Lambda (\tau+r/u)}{1 + \Lambda^2 \left(\tau+r/u\right)^2} 
+ \frac{\Lambda(\tau-r/u)}{1 + \Lambda^2 \left({\tau-r/u}\right)^2}\right]
\end{eqnarray}
whereas
\begin{eqnarray}
&&e^{-\frac{\gamma^2}{2}\langle \left[\phi_+(R+r/2,T_m+\tau/2)-\phi_-(R-r/2,T_m-\tau/2)\right]^2\rangle}
= \nonumber \\
&&\left[\frac{1}{\sqrt{1 + \Lambda^2 (\tau+r/u)^2}}
\frac{1}{\sqrt{1 + \Lambda^2(\tau-r/u)^2}}\right]^{K_{neq}}
\nonumber \\
&& \times \left[\frac{\sqrt{1 + \Lambda^2 \{2(T_m+\tau/2)\}^2}}{\sqrt{1 + \Lambda^2 (2T_m+r/u)^2}}
\frac{\sqrt{1 + \Lambda^2 \{2 (T_m-\tau/2)\}^2}}{\sqrt{1 + \Lambda^2 (2T_m-r/u)^2}}\right]^{K_{tr}}\nonumber \\
&&\times e^{-i K_{eq}\left[
\tan^{-1}\left(\Lambda(\tau+r/u)\right)
+ \tan^{-1}\left(\Lambda (\tau-r/u)\right)
\right]}
\end{eqnarray}
and ${\rm Re}[A] = (A+A^*)/2, {\rm Im}[A] = (A-A^*)/(2i)$.

Collecting all terms we find,
\begin{eqnarray}
&&\delta S^<_g = g\Lambda^2 \int_{-\infty}^{\infty}d(R/u) \int_0^{t} dT_m
\left[\cos\gamma\phi_-^<(R,T_m) \right. \nonumber \\
&&\left. - \cos\gamma\phi_+^<(R,T_m)\right]
e^{-\frac{\gamma^2}{4}\langle\left[\phi_{cl}^>(T_m)\right]^2\rangle}\\
&&\delta S^<_0 = \frac{g^2\gamma^2}{2}\frac{d\Lambda}{\Lambda}
\int_{-\infty}^{\infty} d(R/u)\int_0^{t/\sqrt{2}} dT_m \nonumber \\
&&\left[ -I_R(T_m)
\left(\partial_{R/u} \phi_{cl}^<\right) \left(\partial_{R/u} \phi_{q}^<\right) \right. \nonumber \\
&&\left. - I_{T_m}(T_m)
\left(\partial_{T_m} \phi_{cl}^<\right) \left(\partial_{T_m} \phi_{q}^<\right)\right]\\
&& \delta S_{T_{eff}}^{<}= \frac{ig^2\gamma^2\Lambda^2}{2}\frac{d\Lambda}{\Lambda}
\int_{-\infty}^{\infty} d(R/u)\int_0^{t/\sqrt{2}} dT_m\nonumber \\ 
&&\left(\phi_q^{<}\right)^2  I_{T_{eff}}(T_m)\\
&& \delta S_{\eta}^{<} = -\frac{g^2\gamma^2\Lambda}{2}\frac{d\Lambda}{\Lambda}
\int_{-\infty}^{\infty} d(R/u)\int_0^{t/\sqrt{2}} dT_m \nonumber \\
&&\phi_q^<\left[\partial_{T_m}\phi_{cl}^<\right] I_{\eta}(T_m)
\end{eqnarray}
where
\begin{eqnarray}
&&I_R(T_m) = -\Lambda^4\int_{-\infty}^{\infty} d(r/u) \int_{0}^{2T_m} d\tau (r/u)^2 \nonumber \\
&&{\rm Im}\left[e^{-\frac{\gamma^2}{2}\langle \left[\phi_+(R+r/2,T_m+\tau/2)-\phi_-(R-r/2,T_m-\tau/2)\right]^2\rangle}\right. 
\nonumber\\
&&\left. \times F(r,T_m,\tau)\right]\\
&&I_{T_m}(T_m) = -\Lambda^4\int_{-\infty}^{\infty} d(r/u) \int_{0}^{2T_m} d\tau \tau^2 \nonumber \\
&&{\rm Im}\left[e^{-\frac{\gamma^2}{2}\langle \left[\phi_+(R+r/2,T_m+\tau/2)-\phi_-(R-r/2,T_m-\tau/2)\right]^2\rangle}\right.
\nonumber \\
&&\left. \times F(r,T_m,\tau)\right]\\
&&I_{T_{eff}}(T_m)= \Lambda^2 \int_{-\infty}^{\infty} d(r/u) \int_{-2T_m}^{2T_m} d\tau\nonumber \\  
&&{\rm Re}\left[e^{-\frac{\gamma^2}{2}\langle \left[\phi_+(R+r/2,T_m+\tau/2)-\phi_-(R-r/2,T_m-\tau/2)\right]^2\rangle}\right. 
\nonumber \\
&&\left. \times F(r,T_m,\tau)\right]\\
&&I_{\eta}(T_m)= -\Lambda^3\int_{-\infty}^{\infty} d(r/u) \int_{-2T_m}^{2T_m} d\tau  \\
&&\tau
{\rm Im}\left[e^{-\frac{\gamma^2}{2}\langle \left[\phi_+(R+r/2,T_m+\tau/2)-\phi_-(R-r/2,T_m-\tau/2)\right]^2\rangle}\right. \nonumber \\
&&\left. \times F(r,T_m,\tau)\right]\\
&&I_K = I_R - I_{T_m}\\
&&I_u = I_R + I_{T_m}
\end{eqnarray}

At the next step we rescale the cut-off back to the original value of $\Lambda$, and in the
process rescale position and time to $R,T_m \rightarrow \frac{\Lambda}{\Lambda^{\prime}}(R,T_m)$, where $\Lambda^{\prime}=\Lambda-d\Lambda$.
This rescaling is not necessary in expressions for $\delta S^<_{0,T_{eff},\eta}$ as they are already
of ${\cal O}\left(\frac{d\Lambda}{\Lambda}\right)$. We also express everything in dimensionless units of 
$\bar{R} = \Lambda R/u, {\bar T} = T \Lambda$.
Thus to summarize one obtains, 
\begin{eqnarray}
S_0^{<} &&= \int_{-\infty}^{\infty} d\bar{R} \int_0^{\frac{t}{\sqrt{2}}\Lambda \left(\frac{\Lambda^{\prime}}
{\Lambda}\right)} d
\bar{T}_m
\frac{1}{2\pi K}\left[\phi_q^<\left(\partial_{\bar{R}}^2-\partial_{\bar{T}_m}^2\right)\phi_{cl}^<\right. \nonumber \\
&&\left. +\phi_{cl}^<\left(\partial_{\bar{R}}^2-\partial_{\bar{T}_m}^2\right)\phi_{q}^<
+ \frac{\delta u}{u}\phi_q^<\left(\partial_{\bar{R}}^2+\partial_{\bar{T}_m}^2\right)\phi_{cl}^<
\right. \nonumber \\
&&\left. 
+ \frac{\delta u}{u}\phi_{cl}^<\left(\partial_{\bar{R}}^2+\partial_{\bar{T}_m}^2\right)\phi_{q}^<
- 2\eta\left(\frac{\Lambda}{\Lambda^{\prime}}\right)
\phi_q^<\partial_{\bar{T}_m}\phi_{cl}^< \right. \nonumber \\
&&\left. +i 4 \eta T_{eff}\left(\frac{\Lambda}{\Lambda^{\prime}}\right)^2
\left(\phi_q^<\right)^2\right]
\end{eqnarray}
and
\begin{eqnarray}
&&\delta S^<_g = g \left(\frac{\Lambda}{\Lambda^{\prime}}\right)^2\int_{-\infty}^{\infty}d\bar{R} 
\int_0^{t\Lambda \left(\frac{\Lambda^{\prime}}{\Lambda}\right)} d\bar{T}_m
\left[\cos\gamma\phi_-^<(R,T_m) \right. \nonumber \\
&&\left. - \cos\gamma\phi_+^<(R,T_m)\right]
e^{-\frac{\gamma^2}{4}\langle\left[\phi_{cl}^>(T_m)\right]^2\rangle}\\
&&\delta S^<_0 = \frac{g^2\gamma^2}{2}\frac{d\Lambda}{\Lambda}
\int_{-\infty}^{\infty} d\bar{R}\int_0^{t\Lambda/\sqrt{2}} d\bar{T}_m \nonumber \\
&&\left[ -I_R(T_m)
\left(\partial_{\bar{R}} \phi_{cl}^<\right) \left(\partial_{\bar{R}} \phi_{q}^<\right) - I_{T_m}(T_m)
\left(\partial_{\bar{T}_m} \phi_{cl}^<\right) \left(\partial_{\bar{T}_m} \phi_{q}^<\right)\right]
\nonumber\\\\
&& \delta S_{T_{eff}}^{<}= \frac{ig^2\gamma^2}{2}\frac{d\Lambda}{\Lambda}
\int_{-\infty}^{\infty} d\bar{R}\int_0^{t\Lambda/\sqrt{2}} d\bar{T}_m 
\left(\phi_q^{<}\right)^2 I_{T_{eff}}(T_m)\nonumber \\\\
&& \delta S_{\eta}^{<} = -\frac{g^2\gamma^2}{2}\frac{d\Lambda}{\Lambda}
\int_{-\infty}^{\infty} d\bar{R}\int_0^{t\Lambda/\sqrt{2}} d\bar{T}_m 
\phi_q^<\left[\partial_{\bar{T}_m}\phi_{cl}^<\right] I_{\eta}(T_m)\nonumber \\
\end{eqnarray}
$\delta S_{\eta}$ represents dissipation of the long wavelength modes due to the integrated out high momentum modes, and 
appears as a term proportional to $\phi_q \partial_{T_{m}}\phi_{cl}$, with the strength of the dissipation 
$\eta(T_m)$ depending on the time after the quench. $\delta S_{T_{eff}}$ represents terms that are proportional to $\phi_q^2$
and represents the noise on the long wavelength modes due to the integrated out modes. We denote the strength of this
noise as $\eta(T_m)T_{eff}(T_m)$ because in classical systems, the ratio of the noise and the dissipation strength gives a temperature.
Here too, this allows us to define a time-dependent temperature. 
Defining $\ln{l}$ such that
\begin{eqnarray}
\frac{\Lambda}{\Lambda^{\prime}} = e^{d\ln(l)};
|\frac{d\Lambda}{\Lambda}| = \frac{\Lambda-\Lambda^{\prime}}{\Lambda}=d\ln(l)
\end{eqnarray}
Therefore the RG equations in terms of dimensionless variables such as
$T_m \rightarrow T_m \Lambda, \eta \rightarrow \eta/\Lambda, T_{eff}\rightarrow T_{eff}/\Lambda$
are~\cite{Mitra12b}
\begin{eqnarray}
\frac{dg}{d\ln{l}} = g\left[2-\left(K_{neq} + \frac{K_{tr}}{1+4 T_m^2}\right)\right]
\label{rg1a}\\
\frac{dK^{-1}}{d\ln{l}} = \frac{\pi g^2 \gamma^2}{4}I_K(T_m)\label{rg2a}\\
\frac{1}{Ku}\frac{du}{d\ln{l}} = \frac{\pi g^2 \gamma^2}{4}I_u(T_m)\label{rg3a}\\
\frac{d\eta}{d\ln{l}} = \eta + \frac{\pi g^2\gamma^2 K}{2}I_{\eta}(T_m)\label{rg4a}\\
\frac{d(\eta T_{eff})}{d\ln{l}} = 2 \eta T_{eff} + \frac{\pi g^2\gamma^2 K}{4}I_{T_{eff}}(T_m)
\label{rg5a}\\
\frac{dT_m}{d\ln{l}} = -T_m \label{rg6a}
\end{eqnarray}
Note that the renormalization of the velocity  in Eq.~(\ref{rg3a}) is a minor effect which will be neglected for the rest of the
discussion. In Appendix A, the expression for $I_{K,u,\eta,T_{eff}}$ are presented in dimensionless units.
The physical meaning of the various terms of the $\beta$ function has been discussed in detail in Section~\ref{results}.

\section{Perturbative evaluation of equal time correlation function} \label{pertR}

We now turn to a perturbative evaluation of the correlator $R$ defined in Eq.~(\ref{Rdef}) 
to ${\cal O}(g)$. This will set the stage for the remaining
sections where RG will be used to improve on this result revealing novel nonequilibrium scaling regimes.
Eq.~(\ref{rg1a}) shows that there is a crossover from an intermediate time dynamics where the physics
is determined by the initial wave-function (and hence the initial Luttinger parameter $K_0$) and a long time
time dynamics determined by $K_{neq}$. This can result in a situation where perturbation theory 
in $g$ is violated at intermediate times when $\frac{\gamma^2K_0}{4}<2$ {\sl i.e.}, when $V_{sg}$ is a relevant perturbation in the
initial state. We show below what this implies for $R$. 

Denoting
\begin{eqnarray}
&&R(x_1t,x_2 t) = R^{(0)} + R^{(1)} + \ldots
\end{eqnarray}
where $R^{(i)}$ is the correlator to ${\cal O}(g^i)$, and using results from the previous section, we find
\begin{eqnarray}
&&R^{(0)}(x_1 t,x_2 t) = 2 e^{-\frac{\gamma^2}{8}
\left[\frac{iG_K(11)}{2} + \frac{iG_K(22)}{2}-i G_K(12)\right]}\nonumber \\
&&= 2\left[\left(\frac{1}{\sqrt{1 + \frac{(x_1-x_2)^2}{\alpha^2}}}\right)^{2K_{neq}}\right. \nonumber\\
&&\left. \times\left(\frac{\sqrt{1+ \frac{\left(2ut+x_1-x_2\right)^2}{\alpha^2}}\sqrt{1+ \frac{\left(2ut-x_1+x_2\right)^2}{\alpha^2}}}
{1+\frac{(2ut)^2}{\alpha^2}}
\right)^{-K_{tr}}\right]^{1/4}
\end{eqnarray}
The above gives Eq.~(\ref{R0ll}) in the scaling limit (and setting $t=T_m$).

To next order the equal time correlator is,
\begin{eqnarray}
&&R^{(1)}(x_1t,x_2t) = \frac{4igu}{\alpha^2}\int_{-\infty}^{\infty}dx^{\prime}
\int_0^{t}dt^{\prime}\sum_{c=\pm}sgn(-c)\nonumber \\
&&\langle \cos\left(\gamma\phi_c(x^{\prime}t^{\prime})\right)
 \cos\left(\frac{\gamma}{2}\phi(x_1t)\right) \cos\left(\frac{\gamma}{2}\phi(x_2t)\right)
\rangle\nonumber \\
&&=\frac{igu}{\alpha^2}\int_{-\infty}^{\infty}dx^{\prime}\int_0^{t} dt^{\prime}
\sum_{c=\pm}sgn(-c) \nonumber \\
&&\langle \cos\left(\gamma \phi_c(x^{\prime}t^{\prime})
-\frac{\gamma}{2}\phi(x_1t)-\frac{\gamma}{2}\phi(x_2t)\right)\rangle
\end{eqnarray}

On evaluating the expectation value one finds, 
\begin{eqnarray}
&&R^{(1)}(x_1t,x_2t) = \left(\frac{2gu}{\alpha^2}\right)e^{\frac{\gamma^2}{8}
\left[\frac{iG_K(11)}{2} + \frac{iG_K(22)}{2}-i G_K(12)\right]}\nonumber \\
&&\int_{-\infty}^{\infty}dx^{\prime}\int_0^t dt^{\prime}\nonumber \\
&&\sin\left[\frac{\gamma^2 K}{8}\sum_{\epsilon=\pm}\biggl\{\tan^{-1}\left(\frac{u(t-t^{\prime})+\epsilon (x_1-x^{\prime})}{\alpha}\right)
\right. \nonumber \\
&&\left.+\tan^{-1}\left(\frac{u(t-t^{\prime})+\epsilon(x_2-x^{\prime})}{\alpha}\right)\biggr\}\right]\nonumber \\
&&e^{
-\frac{\gamma^2}{4} \left[
\frac{iG_K(1^{\prime}1^{\prime})}{2} + \frac{iG_K(11)}{2}-i G_K(1^{\prime} 1)
+ \frac{iG_K(1^{\prime}1^{\prime})}{2}
+ \frac{iG_K(22)}{2}-i G_K(1^{\prime} 2)\right]}\nonumber \\
\end{eqnarray}
The above expression shows that at microscopically short times $R^{(1)}(\Lambda t \ll 1) \propto \Lambda t^2$

\begin{figure}
\centering
\includegraphics[totalheight=5.5cm]{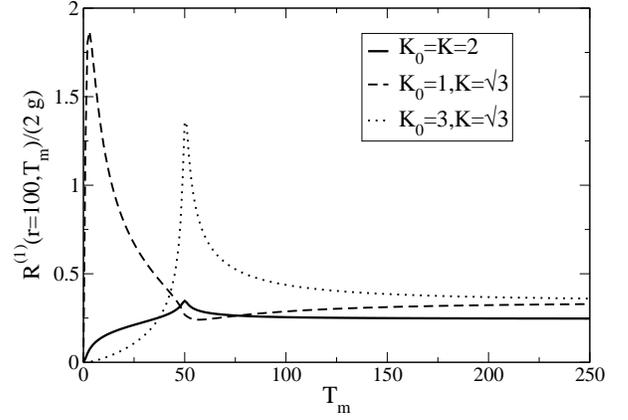}
\caption{The first order correction to the equal time correlation function $R^{(1)}(r=100,T_{m})$ as a function of the time $T_m$
after the quench. Length and time are measured in units of the cut-off $\Lambda$.
Three different quenches are considered: a pure lattice quench  where $K_0=K=2$ (solid line), lattice and
interaction quench corresponding to $K_0=1,K=\sqrt{3}$ (dashed line) and $K_0=3,K=\sqrt{3}$ (dotted line). 
The light-cone is at $T_{m}=r/2=50$.
We choose $\gamma=2$ and all quenches are to the critical point $K_{neq}=2$.}
\label{figR1}
\end{figure}

The behavior of $R^{(1)}$ for several different quench protocols is shown in Fig.~\ref{figR1}. Since the initial state
is gapless, the correlator is non-zero outside the light-cone ($T_{m}<r/2$). Moreover, for a quench where the initial state
is such that the cosine potential is relevant ($\frac{\gamma^2K_0}{4}<2$, dashed line in Fig.~\ref{figR1}), $R$
can be parametrically large at these initial
times outside the light-cone indicating that at these initial times perturbation theory in $g$ is not valid. When the 
initial state is one where the potential is irrelevant or marginally irrelevant ($\frac{\gamma^2K_0}{4}\geq 2$, solid and dotted 
line in Fig.~\ref{figR1}), the correlator is
most enhanced on the light-cone ($r=2T_{m}$). For all these cases, within the light-cone 
($T_{m}>r/2$) the correlator reaches a steady-state. 

We now turn to an RG treatment for the correlator where a $CS$-like equation will be derived.  

\section{Derivation of a Callan-Symanzik-like differential equation out of equilibrium} \label{CSder}

In this section we carry out RG improved perturbation theory and derive a CS-like differential equation for the equal-time correlation function
$R$. We split the fields $\phi$ into slow and fast fields, and integrate over the fast fields.
In doing so, the leading order ($g=0$) correlator $R^{(0)}$ is found to be
\begin{eqnarray}
R^{(0)}(x_1t,x_2t) = R_{<}^0(x_1t,x_2t)
e^{-\frac{d\Lambda}{\Lambda}\gamma_{an,0}(x_1t,x_2t)
\label{R0aa}
}
\end{eqnarray}
where $R_<$ is the correlator for the slow fields and $R$ is the correlator for all the fields, and  
\begin{eqnarray}
&&\gamma_{an,0}(x_1t,x_2t) = \frac{1}{2}\left[K_{neq}\frac{(x_1-x_2)^2\Lambda^2/u^2}
{1+(x_1-x_2)^2\Lambda^2/u^2} \right. \nonumber \\
&&\left. + \frac{K_{tr}}{1+(2t\Lambda)^2}
\right.\nonumber \\
&&\left. -\frac{K_{tr}}{2}\biggl\{\sum_{\epsilon=\pm}\frac{1}{1+\Lambda^2(2t+\epsilon(x_1-x_2)/u)^2}\biggr\}
\right]
\end{eqnarray}
Then we rescale the cut-off, and also rescale position and time. These transformations do not change the above 
expression.
Expanding to ${\cal O}\left(\frac{d\Lambda}{\Lambda}\right)$, and noting that
\begin{eqnarray}
\!\!\Lambda\frac{R^{(0)}(\Lambda)-R^{(0)}_{<}(\Lambda-d\Lambda)}{d\Lambda} = \Lambda \frac{\partial R^{(0)}}{\partial \Lambda}
=\!\!-\frac{\partial R^{(0)}}{\partial\ln{l}}
\end{eqnarray}
we obtain the following CS-like equation to leading order,
\begin{eqnarray}
&&\left[\frac{\partial}{\partial\ln{l}} - \gamma_{an,0}(x_1t,x_2t)\right]R^{(0)}\left(\frac{\Lambda_0}{l}\right) = 0\label{cz0}
\end{eqnarray}
where $\Lambda_0$ is the bare cut-off.

Some limiting expressions for $\gamma_{an,0}$ are as follows:
At long distances and microscopically short times,
\begin{eqnarray}
&&\gamma_{an,0}(|x_1-x_2|\Lambda \gg 1, t \Lambda \ll 1) = \frac{1}{2}\left(K_{neq} + K_{tr}\right) = 
\frac{\gamma^2K_0}{8}\nonumber \\
\end{eqnarray}
At long distances and times and inside the light-cone,
\begin{eqnarray}
&&\gamma_{an,0}(|x_1-x_2|\Lambda \gg 1, t \Lambda \gg 1, 2 t \gg |x_1-x_2|) = \frac{K_{neq}}{2} \nonumber \\ 
\end{eqnarray}
At long distances and times, and outside the light-cone,
\begin{eqnarray}
&&\gamma_{an,0}(|x_1-x_2|\Lambda \gg 1, t \Lambda \gg 1, 2 t \ll |x_1-x_2|) = \frac{K_{neq}}{2} \nonumber \\ 
\end{eqnarray}
At long distances and times, and on the light-cone,
\begin{eqnarray}
&&\gamma_{an,0}(|x_1-x_2|\Lambda \gg 1, t \Lambda \gg 1, 2 t = |x_1-x_2|)  
\nonumber \\
&& =\frac{1}{2}
\left(K_{neq}-\frac{K_{tr}}{2}\right)
\end{eqnarray}

The correlator at next order, $R^{(1)}$, may also be evaluated by splitting it into slow and fast fields as follows
\begin{eqnarray}
&&R^{(1)}(x_1t,x_2t) = 4 {\rm Tr}\left[e^{i(S_0^<+S_0^>)}i S_{sg}(\phi_<+\phi_>)\right. \nonumber \\
&&\left. \times \cos\left(\frac{\gamma \phi^<(x_1t)+\gamma \phi^>(x_1t)}{2}\right)\right. \nonumber \\
&&\left. \times \cos\left(\frac{\gamma \phi^<(x_2t)+\gamma \phi^>(x_2t)}{2}\right)
\right]\nonumber \\
&&=\frac{igu}{\alpha^2}\int_{-\infty}^{\infty}dx^{\prime}\int_0^t dt^{\prime}
\sum_{c=\pm}sgn(-c)  \nonumber \\
&&\langle \cos\left(\gamma \phi_c^<(x^{\prime}t^{\prime})
-\frac{\gamma}{2}\phi^<(x_1t)-\frac{\gamma}{2}\phi^<(x_2t)\right)\rangle\nonumber \\
&&\times \langle\cos\left(\gamma \phi_c^>(x^{\prime}t^{\prime})
-\frac{\gamma}{2}\phi^>(x_1t)-\frac{\gamma}{2}\phi^>(x_2t)\right)\rangle
\end{eqnarray}
On integrating out the fast fields, we obtain,
\begin{eqnarray}
&&R^{(1)}(x_1t,x_2t)= \frac{igu}{\alpha^2} e^{\frac{d\Lambda}{\Lambda}\gamma_{an,0}(x_1,x_2,t)}
\int_{-\infty}^{\infty}dx^{\prime}\int_0^t dt^{\prime}\nonumber \\
&&\sum_{c=\pm}sgn(-c) 
e^{-\frac{d\Lambda}{\Lambda}\delta(x_1,x_2,t,x^{\prime},
t^{\prime})}\nonumber \\
&&\langle \cos\left(\gamma \phi_c^<(x^{\prime}t^{\prime})
-\frac{\gamma}{2}\phi^<(x_1t)-\frac{\gamma}{2}\phi^<(x_2t)\right)\rangle
\end{eqnarray}
where
\begin{eqnarray}
&&\delta(x_1,x_2,t,x^{\prime},t^{\prime}) = 
\left[2K_{neq} + \frac{K_{tr}}{1+(2t^{\prime}\Lambda)^2} + \frac{K_{tr}}{1+(2t\Lambda)^2}
\right.\nonumber \\
&&\left. -\frac{K_{neq}}{2}\sum_{\epsilon=\pm}\biggl\{\frac{1}{1+\Lambda^2(t-t^{\prime}+\epsilon(x_1-x^{\prime})/u)^2}
\biggr\}\right. \nonumber \\
&&\left. -\frac{K_{tr}}{2}\sum_{\epsilon=\pm}\biggl\{\frac{1}{1+\Lambda^2(t+t^{\prime}+\epsilon(x_1-x^{\prime})/u)^2}
\biggr\}\right. \nonumber \\
&&\left.-\frac{K_{neq}}{2}\sum_{\epsilon=\pm}\biggl\{\frac{1}{1+\Lambda^2(t-t^{\prime}+\epsilon(x_2-x^{\prime})/u)^2}
\biggr\}\right. \nonumber \\
&&\left. -\frac{K_{tr}}{2}\sum_{\epsilon=\pm}\biggl\{\frac{1}{1+\Lambda^2(t+t^{\prime}+\epsilon(x_2-x^{\prime})/u)^2}
\biggr\}\right. \nonumber \\
&&\left.-i c\frac{K_{eq}}{2} \sum_{\epsilon=\pm}
\biggl\{\frac{\Lambda(t-t^{\prime}+\epsilon(x_1-x^{\prime})/u)}{1+\Lambda^2(t-t^{\prime}+\epsilon(x_1-x^{\prime})/u)^2}
\biggr\}
\right. 
\nonumber \\
&&\left. 
-i c \frac{K_{eq}}{2} 
\sum_{\epsilon=\pm}\biggl\{\frac{\Lambda(t-t^{\prime}+\epsilon(x_2-x^{\prime})/u)}{1+\Lambda^2(t-t^{\prime}+\epsilon(x_2-x^{\prime})/u)^2}
\biggr\}
\right]
\end{eqnarray}
Rescaling the cut-off and correspondingly the position and time, we obtain the following differential equation
\begin{eqnarray}
&&R^{(1)}(x_1-x_2,t) = R_<^{(1)}(x_1-x_2,t)
\left[1 + 2\frac{d\Lambda}{\Lambda} - 2 K_{neq} \frac{d\Lambda}{\Lambda} \right. \nonumber \\
&&\left. + \gamma_{an,0}(x_1-x_2,t)
\frac{d\Lambda}{\Lambda}\right] \nonumber \\
&&+ 2 g \pi I_C(x_1,x_2,t) \frac{d\Lambda}{\Lambda}R_<^0(x_1-x_2,t)\label{r1}
\end{eqnarray}
Rewriting the zero-order term Eq.~(\ref{R0aa}) for convenience,
\begin{eqnarray}
R^{(0)}(x_1-x_2,t) &&= R^{(0)}_<(x_1-x_2,t)\left[1\right. \nonumber \\
&&\left. -\frac{d\Lambda}{\Lambda}\gamma_{an,0}(x_1-x_2,t)\right]\label{r0}
\end{eqnarray}
and combining Eq.~(\ref{r1}),~(\ref{r0}), we get
\begin{eqnarray}
&&R = R^{(0)}_<\left[1-\frac{d\Lambda}{\Lambda}\gamma_{an,0}(x_1-x_2,t)\right. \nonumber \\
&&\left. + \frac{d\Lambda}{\Lambda} 2g\pi I_C(x_1-x_2,t)\right] \nonumber \\
&&+ R^{(1)}_<
\left[1 + (2-K_{neq})\frac{d\Lambda}{\Lambda} \right. \nonumber \\
&&\left. + \biggl\{-K_{neq}+\gamma_{an,0}(x_1-x_2,t)\biggr\}
\frac{d\Lambda}{\Lambda}\right]
\end{eqnarray}
The term in the first square brackets ($\gamma_{an,0}-2\pi gI_C$) is the anomalous scaling dimension of the correlator $R$.
The $(2-K_{neq})d\Lambda/\Lambda$ term in the second square bracket is simply the renormalization of $g$ which has been discussed before
in the context of the $\beta$-function. 

To next order in the cosine potential, we obtain the ${\cal O}(g^2)$ terms of the $\beta$-function whose derivation is already discussed
in Section~\ref{beta}. Thus to ${\cal O}(g^2)$ the following CS differential-equation for $R$ is obtained,
\begin{eqnarray}
&&\left[\frac{\partial}{\partial \ln{l}} + \beta(g_i) \frac{\partial }{\partial g_i}- \gamma_{an,0} + 2\pi g I_C\right]\nonumber \\
&&\times R\left[\frac{r\Lambda_0}{l}, 
\frac{\Lambda_0T_{m0}}{l},g_i(l)\right]=0
\label{cz1}
\end{eqnarray}
where $g_i = g,K, \eta,T_{eff}$, and $\Lambda_0, T_{m0}, g_{i0}$ denote bare values. Moreover $I_C$ is given by 
\begin{widetext}
\begin{eqnarray}
&&I_C(r=x_1-x_2,t,K_{neq},K_{eq}) = -\left(\frac{1}{2\pi}\right)e^{K_{neq}\ln\sqrt{1+(x_1-x_2)^2}}\nonumber \\
&&\int_{-\infty}^{\infty}dx^{\prime}\int_0^t dt^{\prime}
\sin\left[\frac{K_{eq}}{2} \sum_{\epsilon=\pm}\biggl\{\tan^{-1}\left(t^{\prime}+\epsilon (x^{\prime}-x_1)\right)
+ \tan^{-1}\left(t^{\prime}+\epsilon(x^{\prime}-x_2)\right)
\biggr\}
\right]\nonumber \\
&&\times e^{-\frac{K_{neq}}{2}\sum_{\epsilon=\pm}\left[\ln\sqrt{1+\left(t^{\prime}+\epsilon(x^{\prime}-x_1)\right)^2}+ 
\ln\sqrt{1+\left(t^{\prime}+\epsilon(x^{\prime}-x_2)\right)^2} 
\right]}\nonumber \\
&&\times e^{-\frac{K_{tr}}{2}\left[\ln\sqrt{\frac{1+\left(2t-t^{\prime}+x^{\prime}-x_1\right)^2}{1+4t^2}}+ 
\ln\sqrt{\frac{1+\left(2t - t^{\prime}-(x^{\prime}-x_1)\right)^2}{1+4(t-t^{\prime})^2}} 
+ \ln\sqrt{\frac{1+\left(2t-t^{\prime}+x^{\prime}-x_2\right)^2}{1+4t^2}} 
+\ln\sqrt{\frac{1+\left(2t-t^{\prime}-(x^{\prime}-x_2)\right)^2}{1+4(t-t^{\prime})^2}}
\right]}\nonumber \\
&&\times\left[\frac{K_{tr}}{1+4t^2} +\frac{K_{tr}}{1+4 (t-t^{\prime})^2}
-\frac{K_{neq}}{2}\sum_{\epsilon=\pm}\biggl\{
\frac{1}{1+\left(t^{\prime}+ \epsilon(x^{\prime}-x_1)\right)^2}
+\frac{1}{1+\left(t^{\prime}+ \epsilon(x^{\prime}-x_2)\right)^2}
\biggr\}\right. \nonumber \\
&&\left. -\frac{K_{tr}}{2}\sum_{\epsilon=\pm}\biggl\{
\frac{1}{1+(2t-t^{\prime}+\epsilon(x^{\prime}-x_1))^2} 
+ \frac{1}{1+(2t-t^{\prime}+\epsilon(x^{\prime}-x_2))^2}\biggr\}
\right]\label{ic1}\\
&&+\frac{1}{2\pi}
e^{K_{neq}\ln\sqrt{1+(x_1-x_2)^2}}
\int_{-\infty}^{\infty}dx^{\prime}\int_0^{t}dt^{\prime}
\cos\left[\frac{K_{eq}}{2}\sum_{\epsilon=\pm}\biggl\{ \tan^{-1}\left(t^{\prime}+\epsilon(x^{\prime}-x_1)\right)
+ \tan^{-1}\left(t^{\prime}+\epsilon(x^{\prime}-x_2)\right)
\biggr\}
\right]\nonumber \\
&&\times e^{-\frac{K_{neq}}{2}\sum_{\epsilon=\pm}\left[\ln\sqrt{1+\left(t^{\prime}+\epsilon(x^{\prime}-x_1)\right)^2}+ 
\ln\sqrt{1+\left(t^{\prime}+\epsilon(x^{\prime}-x_2)\right)^2} 
\right]}\nonumber \\
&&\times e^{-\frac{K_{tr}}{2}\left[\ln\sqrt{\frac{1+\left(2t-t^{\prime}+x^{\prime}-x_1\right)^2}{1+4t^2}}+ 
\ln\sqrt{\frac{1+\left(2t - t^{\prime}-(x^{\prime}-x_1)\right)^2}{1+4(t-t^{\prime})^2}} 
+ \ln\sqrt{\frac{1+\left(2t-t^{\prime}+x^{\prime}-x_2\right)^2}{1+4t^2}} 
+\ln\sqrt{\frac{1+\left(2t-t^{\prime}-(x^{\prime}-x_2)\right)^2}{1+4(t-t^{\prime})^2}}
\right]} \nonumber\\
&&\times \frac{K_{eq}}{2}\sum_{\epsilon=\pm}
\left[
\frac{\left(t^{\prime}+\epsilon(x^{\prime}-x_1)\right)}{1+\left(t^{\prime}+ \epsilon(x^{\prime}-x_1)\right)^2}
+\frac{\left(t^{\prime}+\epsilon(x^{\prime}-x_2)\right)}{1+\left(t^{\prime}+ \epsilon(x^{\prime}-x_2)\right)^2}
\right]\label{ic2}
\end{eqnarray}
\end{widetext}

The usual CS equation encountered in equilibrium for
the Heisenberg chain~\cite{Singh89, Affleck98, Barzykin99} may be obtained from above by setting the time
after the quench $t=\infty$, and by setting $K_0=K$ and by also  
being at the critical point $K_{eq}=2$. Here,
noting that  $-{\rm Im}\left[e^{-i\tan^{-1}x}\frac{1}{1+i x}\right]=\frac{\sin\left(\tan^{-1}x\right)}{1+x^2}
+ \frac{x}{1+x^2}\cos\left(\tan^{-1}x\right)$,
one may write
\begin{eqnarray}
&&I_C = -\frac{\left(1+(x_1-x_2)^2\right)}{2\pi} {\rm Im}\left[
\int_{-\infty}^{\infty}dx^{\prime}\int_0^{\infty}dt^{\prime}\right. \nonumber \\
&&\left. e^{-\sum_{\epsilon=\pm}\ln\left(1+i(t^{\prime}+\epsilon(x^{\prime}-x_1))\right)
+\ln\left(1+i(t^{\prime}+\epsilon(x^{\prime}-x_2))\right)}
\right. \nonumber \\
&&\left. \sum_{\epsilon=\pm}\biggl\{
\frac{1}{1+i(t^{\prime}+\epsilon(x^{\prime}-x_1))}
+ \frac{1}{1+i(t^{\prime}+\epsilon(x^{\prime}-x_2))}\biggr\}
\right]\nonumber \\
\end{eqnarray}
The above integral may be evaluated to give, 
\begin{eqnarray}
I_C 
= \frac{(x_1-x_2)^2+1}{4 + (x_1-x_2)^2}
\end{eqnarray}
Thus for $|x_1-x_2|\gg 1$,
\begin{eqnarray}
I_C\left(K_0=K,K_{eq}=2,t=\infty\right) = 1
\end{eqnarray}
In the next two sections we will solve the 
CS-equation~(\ref{cz1}) for two cases, one is when the Luttinger liquid interaction parameter is held fixed, but the cosine or lattice
potential is suddenly switched on, and the second is when the Luttinger parameter is changed at the same time as when the lattice
potential is switched on.

\section{Correlation function for the pure lattice quench}\label{lattice}
In the previous section we showed that in order to determine the correlation function at 
spatial separation $r$ and a time $T_{m0}$ after a quench, we need to solve 
\begin{eqnarray}
&&\left[\frac{\partial}{\partial \ln{l}} + \beta(g_i) \frac{\partial }{\partial g_i}- \gamma_{an,0} + 2\pi g I_C\right]\nonumber \\
&&\times R\left[\frac{r\Lambda_0}{l}, 
\frac{\Lambda_0T_{m0}}{l},g_i(l)\right]=0
\label{cz2}
\end{eqnarray}
where $g_i = g,K,\eta, T_{eff}$, and $\Lambda_0, g_{i0}$ denote bare values.
In this section we will solve Eq.~(\ref{cz2}) for the lattice quench $K_0=K$ and near the critical point 
$K_{eq} = 2 + \delta, \forall\,\, 0<\delta\ll 1$.
For this case, $\gamma_{an,0}$ has the following limiting forms 
\begin{eqnarray}
&&\gamma_{an,0}(r\Lambda \gg 1, T_m \Lambda \ll 1) = \frac{1}{2}\left(K_{neq} + K_{tr}\right) \nonumber \\
&&=1 + \frac{\delta}{2}\\
&&\gamma_{an,0}(r\Lambda \gg 1, T_m \Lambda \gg 1, 2 T_m \gg r) = \frac{K_{neq}}{2}\nonumber \\
&&=1 +\frac{\delta}{2} \\
&&\gamma_{an,0}(r\Lambda \gg 1, T_m \Lambda \gg 1, 2 T_m \ll r) = \frac{K_{neq}}{2}\nonumber \\
&&=1 + \frac{\delta}{2}\\
&&\gamma_{an,0}(r\Lambda \gg 1, T_m \Lambda \gg 1, 2 T_m = r) \nonumber \\
&&= \frac{1}{2}
\left(K_{neq}-\frac{K_{tr}}{2}\right)
=1 + \frac{\delta}{2}
\end{eqnarray}
Moreover for $K_0=K$ and in the vicinity of the critical point, the full expression for $I_C$ in Eqs.~(\ref{ic1}),~(\ref{ic2}) reduces to 
\begin{eqnarray}
&&I_C(r,T_m) = \frac{r^2+1}{4 + r^2} \nonumber \\
&&- \frac{r^2+1}{2\pi}{\rm Im}\left[i \int_{-\infty}^{\infty}dx
\biggl\{\frac{1}{(1+iT_m)^2 + x^2}\right. \nonumber \\
&&\left. \times 
\frac{1}{(1+iT_m)^2 + (x-r)^2}\biggr\}\right]\nonumber \\
&&=\frac{r^2+1}{4 + r^2}  \nonumber \\
&&-\frac{1+r^2}{1 + T_m^2}
\left[\frac{r^2-4T_m^2 + 4 -8T_m^2}{\biggl\{r^2 - 4 T_m^2 + 4\biggr\}^2 + 64 T_m^2}\right]\nonumber \\
\end{eqnarray}
$I_C$ has the following different limits already discussed in Section~\ref{results},
\begin{eqnarray}
&&I_C(r\gg 1,T_m\gg 1, 2T_m\gg r) \nonumber \\
&&=1 + {\cal O}\left(\frac{1}{r^2},\frac{r^2}{T_m^4}\right)\\
&&I_C(r\gg 1,T_m\gg 1, 2T_m \ll r) \nonumber \\
&&=1 + {\cal O}\left(\frac{1}{T_m^2},\frac{1}{r^2}\right)\\
&&I_C(r\gg 1,T_m\gg 1, 2T_m  = r) \nonumber \\
&&= \frac{3}{2} + {\cal O}\left(\frac{1}{r^2}\right)
\end{eqnarray}
In addition the $\beta$-function for a pure lattice quench and using $K_{neq}= K_{eq}= 2 + \delta$ is
(neglecting velocity renormalization),
\begin{eqnarray}
\frac{dg}{d\ln{l}} = -g\delta
\label{rg1lc}\\
\frac{d\delta}{d\ln{l}} = -4 \pi g^2 I_K(T_m)\label{rg2lc}\\
\frac{d\eta}{d\ln{l}} = \eta + 4 \pi g^2I_{\eta}(T_m)\label{rg4lc}\\
\frac{d(\eta T_{eff})}{d\ln{l}} = 2 \eta T_{eff} + 2\pi g^2I_{T_{eff}}(T_m)
\label{rg5lc}\\
\frac{dT_m}{d\ln{l}} = -T_m \label{rg6lc}
\end{eqnarray}
where
\begin{eqnarray}
&&I_K(T_m, K_{neq}=K_{eq}=2) \nonumber \\
&&= \pi - \frac{\pi}{2}\left[\frac{2+20T_m^2 + 7\times 16 T_m^4}{(1+ 4 T_m^2)^3}\right]
\end{eqnarray}
The above shows that at short times, 
$I_K(T_m\ll 1) = 2\pi T_m^2 + \ldots$, while at long times $I_K$ reaches a steady-state as follows,
$I_K(T_m\gg 1) = \pi \left[1-\frac{7}{8T_m^2} + \ldots\right]$.
Moreover,
\begin{eqnarray}
&&\!\!\!\!I_{\eta}(T_m, K_{neq}=K_{eq}=2)=4\pi \frac{(2 T_m)^3}{\left[1 + (2 T_m)^2\right]^3}\\
&&\!\!\!\!I_{T_{eff}}(T_m, K_{neq}=K_{eq}=2) \!\!= \!\!6\pi T_m \left[\frac{1-\frac{4}{3}T_m^2}{(1+ 4 T_m^2)^3}\right]
\end{eqnarray}
Note that the above expressions show that at long times ($T_{m}\rightarrow \infty$) 
a pure lattice quench does not generate any dissipation and noise, at least to ${\cal O}(g^2)$.
This extends the regime of validity of the prethermalized regime which we defined as the regime where 
time is larger than microscopic
time-scales, but smaller than the steady-state inelastic scattering rate (Fig.~\ref{fig3}). 

Thus neglecting dissipative and thermal effects, and at macroscopically long times,  
the $\beta$-function simplifies considerably to
\begin{eqnarray}
\frac{dg}{d\ln{l}} = -g\delta;\frac{d\delta}{d\ln{l}}=-(2\pi g)^2 \label{beta1a}
\end{eqnarray}
with the anomalous dimension given by 
\begin{eqnarray}
\gamma_{an} = \gamma_{an,0} - 2\pi g I_C= 1 +\frac{\delta}{2} - 2\pi g I_C
\end{eqnarray}
As discussed earlier in this section, $I_C=1$ inside ($2T_{m0}\gg r$) and outside ($2T_{m0}\ll r$) the
light-cone, whereas $I_C=3/2$ for points on the light-cone $r=2T_{m0}$.
We will derive expressions for the correlator for two cases separately. The first case is when the final Hamiltonian is on the 
critical line $\delta=2\pi g$, while the second is when the final Hamiltonian is slightly away
from the critical line $\delta > 2\pi g$.

For the first case, {\sl i.e.}, on the critical line $\delta=2\pi g$, the $\beta$ function further simplifies to
\begin{eqnarray} 
\frac{d\delta}{d\ln{l}} = -\delta^2\label{bet1}
\end{eqnarray}
and the anomalous dimension of the correlator becomes
\begin{eqnarray}
\gamma_{an} = 1 +\frac{\delta}{2} - \delta I_C
\end{eqnarray}
The explicit forms of $I_C$ and $\gamma_{an,0}$ in Eq.~(\ref{gan0}),~(\ref{IC1}) show that 
scaling stops when $l^* = {\rm min}\left(\Lambda_0 r,\Lambda_0T_{m0}\right)$. Thus there
are three interesting cases to consider, one for spatial separations outside the light-cone $r \gg 2 T_{m0}$, 
the second is for spatial separations on the light-cone 
$r = 2T_{m0}$, and the third is for spatial separations within a light-cone $r \ll 2 T_{m0}$. These three cases are shown pictorially in
Fig.~\ref{fig2}, and the corresponding correlators are derived next.

Integrating the the CS-equation upto $l^*$ we obtain
\begin{eqnarray}
&&R\left(\Lambda_0 r,\Lambda_0T_{m0},g_0\right) \nonumber \\
&&= e^{-\int_{g_0}^{g(l^*)} dg^{\prime}\frac{\gamma_{an}(g^{\prime})}{\beta(g^{\prime})}}
R\left(\frac{r\Lambda_0}{l^*}, \frac{T_{m0}\Lambda_0}{l^*},g(l^*)\right)
\end{eqnarray}
Integrating upto $l^*$, the solution of Eq.~(\ref{bet1}) is
\begin{eqnarray}
\frac{1}{\delta(l^*)}-\frac{1}{\delta_0} = \ln\left(l^*\right)
\end{eqnarray}
whereas for evaluating the correlator we need
\begin{eqnarray}
&&\int_{\delta_0}^{\delta(l^*)}d\delta\frac{\gamma_{an}}{\beta} = \ln(l^*) \nonumber \\
&&+ \left(I_C-\frac{1}{2}\right)
\ln\left[\frac{\delta(l^*)}{\delta_0}\right]
\end{eqnarray}

For points outside the light-cone, setting $l^*= 2\Lambda_0 T_{m0}$, the correlation function is found to be
\begin{eqnarray}
&&R\left(\Lambda_0 r,\Lambda_0 T_{m0},g_0, 2T_{m0}\ll r\right)\sim \frac{\sqrt{\ln{T_{m0}}}}{T_{m0}}\nonumber \\
&&\times R\left(\frac{r}{2T_{m0}}\gg 1, T_{m0}\Lambda_0=l^*, g(l^*)\right)
\end{eqnarray}
Since $l^*\gg 1$, $g(l^*)\ll g_0$, so that the short-distance correlator may be evaluated perturbatively in the 
cosine potential. Using the results from Eq.~(\ref{Csd3}) where $R\left(\frac{r}{2T_{m0}}, T_{m0}\Lambda_0=l^*, g=0\right)
~\sim T_{m0}/r$, we find that the correlator outside the light-cone behaves as
\begin{eqnarray}
R\left(\Lambda_0 r,g_0, 2T_{m0}\ll r\right)\sim \frac{\sqrt{\ln{T_{m0}}}}{r}\label{Clcout}
\end{eqnarray}

The second interesting case is the behavior of the correlator inside the light-cone where
$2T_{m0}\gg r$. Here integrating upto
$l^* = \Lambda_0 r$, we get
\begin{eqnarray}
&&R\left(\Lambda_0 r,2T_{m0}\gg r,g_0\right) \nonumber \\
&&= 
e^{-\int_{g_0}^{g(\Lambda_0 r)} dg^{\prime}\frac{\gamma_{an}(g^{\prime})}{\beta(g^{\prime})}}R\left(\frac{r\Lambda_0}{l^*}=1, 2T_{m0}\gg r,
g(l^*)\right)\nonumber \\
\label{czsol2a}
\end{eqnarray}
From Eq.~(\ref{Csd1}), $R\left(l^*=\Lambda_0 r, 2T_{m0}\gg r,g=0\right) = {\cal O}(1)$, while using the result that $I_C=1$ within the light-cone
we obtain
\begin{eqnarray}
R\left(\Lambda_0 r,2T_{m0}\gg r,g_0\right)\sim \frac{\sqrt{\ln{r}}}{r} \label{lcina}
\end{eqnarray}
At these long times, the explicit dependence on the time
after the quench drops out and the correlator reaches a steady-state value. Moreover this result is the
same as in the ground state of the final Hamiltonian. Later when we study an interaction and lattice quench,
we will see that within the light-cone, a steady-state behavior again arises, however the logarithmic corrections
are different those obtained for a system which is in equilibrium, and near the critical point.

The third interesting case is for spatial separations on the light-cone where $r = 2T_{m0}$. Here
integrating up to $l^*=\Lambda_0 r$
\begin{eqnarray}
&&R\left(\Lambda_0 r,2T_{m0}=r,g_0\right) = 
e^{-\int_{g_0}^{g(\Lambda_0 r)} dg^{\prime}\frac{\gamma_{an}(g^{\prime})}{\beta(g^{\prime})}}\nonumber \\
&&\times R\left(\frac{r\Lambda_0}{l^*}=1, 2T_{m0}=l^*,
g(l^*)\right)
\label{czsol3a}
\end{eqnarray}
Using Eq.~(\ref{Csd2}), and the fact that $I_C=3/2$ on the light-cone we find
\begin{eqnarray}
R\left(\Lambda_0 r,2T_{m0}=r,g_0\right) \sim \frac{\ln{r}}{r}\label{lcona}
\end{eqnarray}
On the light-cone ($r=2T_{m0}$)  
the correlator decays somewhat slower than within and outside the light-cone. 

We now turn to the case where $\delta > 2\pi g$. Here the solution of Eq.~(\ref{beta1a}) 
on integrating upto a scale $l^*$ (defining 
$A_1=\sqrt{\delta^2-(2\pi g)^2}$) is,
\begin{eqnarray}
&&\int_{g_0}^{g(l^*)}\frac{\gamma(g^{\prime})}{\beta(g^{\prime})}dg^{\prime}=\nonumber \\
&&\ln{l^*}
+ \frac{1}{2}\ln\left[\frac{\sinh\left(A_1\ln{l^*} + \tanh^{-1}\frac{A_1}{\delta_0}\right)}{\sinh\left(\tanh^{-1}\frac{A_1}{\delta_0}\right)}\right]
\nonumber \\
&&- I_CA_1\ln\left[\frac{\tanh\left(\frac{A_1}{2}\ln{l^*} + \frac{1}{2}
\tanh^{-1}\frac{A_1}{\delta_0}\right)}{\tanh\left(\frac{1}{2}\tanh^{-1}\frac{A_1}{\delta_0}\right)}\right]
\end{eqnarray}
The correlation function therefore is found to be
\begin{eqnarray}
&&R\left(\Lambda_0 r, \Lambda_0 T_{m0}, g_0\right)\simeq \nonumber \\
&&\frac{1}{\left(l^*\right)^{1+A_1/2}}\sqrt{\frac{1}{1-(l^*)^{-2A_1}}}
\left[\frac{1-(l^*)^{-A_1}}{1+(l^*)^{-A_1}}\right]^{I_C}\label{Rgencr}
\end{eqnarray}

For points outside the light-cone we set $l^*=\Lambda_0T_{m0}$ in Eq.~(\ref{Rgencr}), and using Eq.~(\ref{Csd3}) we obtain
\begin{eqnarray}
&&R\left(\Lambda_0r,\Lambda_0T_{m0},g_0, r \gg 2T_{m0}\right)\sim \nonumber \\
&&\left(\frac{1}{r}\right)^{1+\delta/2}\left(T_{m0}\right)^{\frac{\delta}{2}-\frac{\sqrt{\delta^2-(2\pi g)^2}}{2}}
\label{Rolc}
\end{eqnarray}
For points on the light-cone, or inside the light-cone we set $l^* = \Lambda_0 r$ in Eq.~(\ref{Rgencr}) to obtain,
\begin{eqnarray}
&&R\left(\Lambda_0r,\Lambda_0T_{m0},g_0, r \leq  2T_{m0}\right)\sim \frac{1}{r^{1+A_1/2}}
\end{eqnarray}
Note that for small deviations away from the critical point such that $A_1\ln{r}\gg 1$, the correlator on the light-cone
and inside the light-cone have the same leading behavior in position. This is however not the case for quenches to the critical
point where $A_1 \ln{r}\ll 1$. Here Eqs.~(\ref{lcina}) and~(\ref{lcona}) show that 
differences arise even in the leading asymptote. 

\section{Correlation function for the lattice and interaction quench} \label{latticeint}

The $\beta$-function shows that an interaction quench $K_0\neq K$ changes the location of the critical point to $K_{neq}=2$.
Since $K_{neq}>K_{eq}$, this implies that the ground state of the final Hamiltonian can be in the gapped phase, however
the quench results in a highly excited and more delocalized state of bosons. 
In this section we are interested in evaluating the correlator $R$ in the vicinity of this new 
nonequilibrium critical point. Setting $K_{neq}= 2$ implies the following relation between $K_0$ and $K$, 
\begin{eqnarray}
&&K = K_0 \sqrt{\frac{16}{\gamma^2K_0}-1}
\end{eqnarray}
This implies the following,
\begin{eqnarray}
&&K_{eq}= \frac{\gamma^2K_0}{4} \sqrt{\frac{16}{\gamma^2K_0}-1}\label{keqneq}\\
&&K_{tr} = \frac{\gamma^2K_0}{4}-2\label{ktrneq}
\end{eqnarray}

When $K_0\neq K$, dissipative effects are generated. However in the prethermalized regime these effects are still
weak and may be neglected.
Writing $K_{neq}=2+\delta$, the RG equations in the prethermalized regime ($T_m < 1/\eta$) are the following 
\begin{eqnarray}
&&\frac{dg}{d\ln{l}} = -g \delta\\
&&\frac{d\delta}{d\ln{l}} = -\pi g^2 \left(\frac{\gamma^2K_0}{4}\right)^2\left(\frac{16}{\gamma^2K_0}-1\right)^{3/2}\!\!\!\!\!I_K(T_m)\\
&& \frac{dT_m}{d\ln{l}}= -T_m
\end{eqnarray}
above we have used that since $K_{neq}=2 + \delta$, $dK = \frac{4}{\gamma^2}\frac{K_0}{K}d\delta$, and $K_{eq}$ is given by Eq.~(\ref{keqneq})
Moreover,
\begin{eqnarray}
&&I_K(T_m=\infty,K_{neq}=2) = \nonumber \\
&&-4 \int_0^{\infty}du \frac{u}{(1+u^2)^2}\cos\left(K_{eq}\tan^{-1}{u}\right)\nonumber \\
&&\times \underbrace{\int_0^u dv \frac{v}{1+v^2}
\sin\left(K_{eq}\tan^{-1}{v}\right)}_{i_1(u)}\label{ik1}\\
&&-4 \int_0^{\infty}du \frac{u}{1+u^2}\cos\left(K_{eq}\tan^{-1}{u}\right)\nonumber \\
&&\times \underbrace{\int_0^u dv \frac{v}{(1+v^2)^2}\sin\left(K_{eq}\tan^{-1}{v}\right)}_{i_2(u)}\label{log1}\\
&&-2 K_{eq}\int_0^{\infty}du \frac{u}{1+u^2}\cos\left(K_{eq}\tan^{-1}{u}\right)\nonumber \\
&&\times \underbrace{\int_0^u dv \frac{v^2}{(1+v^2)^2}\cos\left(K_{eq}\tan^{-1}{v}\right)}_{i_3(u)}\label{log2}\\
&&+ 2 K_{eq}\int_0^{\infty}du \frac{u^2}{(1+u^2)^2}\sin\left(K_{eq}\tan^{-1}{u}\right)\nonumber \\
&&\times \underbrace{\int_0^u dv \frac{v}{1+v^2}\sin\left(K_{eq}\tan^{-1}{v}\right)}_{i_1(u)}\label{ik4}
\end{eqnarray}
Note that Eqs.~(\ref{log1}),~(\ref{log2}) are logarithmically divergent (such a
divergence does not arise for a pure lattice quench). The reason for this divergence is because
we have used leading order perturbation theory to evaluate the correlators that go into $I_K$, whereas, we should have
used the correlators from the full theory which correspond to a non-zero dissipation $\eta$. Taking this
into account, the logarithmic divergence is
cut-off by ${\rm min}(T_m,1/\eta)$.
Denoting the
well behaved terms in Eqs.~(\ref{ik1}),~(\ref{ik4}) as $\pi c_1$,  and using
$i_2(\infty)= -\frac{1}{K_{eq}^2-4}\sin\left(\frac{\pi K_{eq}}{2}\right)$ and $i_3(\infty)
= \frac{(K_{eq}^2-2)}{K_{eq}(K_{eq}^2-4)}\sin\left(\frac{\pi K_{eq}}{2}\right)$, we may write,
\begin{eqnarray}
&&I_K(T_m,K_{neq}=2) = \nonumber \\
&&\pi\left[c_1 -2 c^{\prime}_2 \sin\left(\frac{\pi K_{eq}}{2}\right)\right. \nonumber \\
&&\left. \times \int_0^{min(T_m,1/\eta)}du \frac{u}{1+u^2}\cos\left(K_{eq}\tan^{-1}{u}\right)\right]\\
&&\simeq\pi\left[c_1 -c_2 \sin\left(\pi K_{eq}\right)\ln\left({\rm min}(T_m,1/\eta)\right)\right]
\end{eqnarray}
where $c_{1,2}$ are ${\cal O}(1)$ and depend on $K_{eq}$ and hence $K_0$. When $K_0=K$, $c_1=1$. 

In what follows we will consider a prethermalized regime where $T_m < 1/\eta$, and also weak
quenches such that $|K_{eq}-K_{neq}|\ln{T_m} \ll 1$.
In this case, $I_K\simeq \pi c_1$ and is independent of time. 
We will solve the RG equations in a regime where the periodic potential is marginally irrelevant
$g_{eff} = g B <\delta$
where
\begin{eqnarray}
B =  \pi \sqrt{c_1}\left(\frac{\gamma^2K_0}{4}\right)\left(\frac{16}{\gamma^2K_0}-1\right)^{3/4}
\end{eqnarray}
The RG flow equations $\frac{dg_{eff}}{d\ln{l}}=-g_{eff}\delta, \frac{d\delta}{d\ln{l}}=-g_{eff}^2$
are characterized by the constant of motion
\begin{eqnarray}
A = \sqrt{\delta^2 - g_{eff}^2}
\end{eqnarray}
in terms of which the solution of the RG equations are,
\begin{eqnarray}
\delta(l) = A \coth\left[A\ln{l} + \tanh^{-1}\left(\frac{A}{\delta_0}\right)\right]\\
g_{eff}(l) =\frac{A}{\sinh\left[A\ln{l}+ \tanh^{-1}\left(\frac{A}{\delta_0}\right) \right]}
\end{eqnarray}

In the vicinity of the nonequilibrium critical point $K_{neq}=2+\delta,\forall\,\, \delta \ll 1$, 
and inside the light-cone, $I_C$ is found to be
\begin{eqnarray}
&&I_C(r\gg 1, T_m \gg 1, 2T_m \gg r)=  \frac{K_{eq}}{2}\nonumber \\
&&+ \left[1-\left(\frac{K_{eq}}{2}\right)^2\right]L(K_{eq})
\label{IClt}
\end{eqnarray}
where
\begin{widetext}
\begin{eqnarray}
&&L(K_{eq}) = \lim_{|r|\gg 1}\left(\frac{r^2}{4\pi} \int_0^{\infty}du \int_{-u}^u dv \sin\left[\frac{K_{eq}}{2}\biggl\{\tan^{-1}u + 
\tan^{-1}v + \tan^{-1}(u+r) + \tan^{-1}(v-r)\biggr\}\right]\right. \nonumber \\
&&\left. \frac{1}{\sqrt{1+u^2}}\frac{1}{\sqrt{1+v^2}}
\frac{1}{\sqrt{1+(u+r)^2}}\frac{1}{\sqrt{1+(v-r)^2}}
\left[\frac{1}{1+u^2}+ \frac{1}{1+v^2}+ \frac{1}{1+(u+r)^2} + \frac{1}{1+(v-r)^2}\right]\right)\\
&& + \left(r \rightarrow - r\right)\nonumber
\end{eqnarray}
\end{widetext}
$L(K_{eq})$ is a universal function of $K_{eq}$ and hence the initial Luttinger parameter. This results in a
universal expression for $I_C$ which is plotted in
the Fig.~\ref{figL} for different initial states. 

The solution of the CS equation is,
\begin{eqnarray}
&&R\left(\Lambda_0 r,\Lambda_0T_{m0},g_0\right) = \nonumber \\
e^{-\int_{g_0}^{g(l)} dg^{\prime}\frac{\gamma_{an}(g^{\prime})}{\beta(g^{\prime})}}
&&R\left(\frac{r\Lambda_0}{l}, \frac{T_{m0}\Lambda_0}{l},g(l)\right)
\end{eqnarray}
We will solve this first for points within the light-cone 
$1/\eta \gg T_{m0}\gg r$. For this case, dissipative effects are neglected and 
the RG equations are integrated upto $l=\Lambda_0 r$ to give,
\begin{eqnarray}
&&R\left(\Lambda_0 r, r \ll T_{m0}\ll 1/\eta,g_0\right) = \nonumber \\
&&e^{-\int_{g_0}^{g(\Lambda_0 r)} dg^{\prime}\frac{\gamma_{an}(g^{\prime})}{\beta(g^{\prime})}}
R\left(\frac{r\Lambda_0}{l}=1, \frac{T_{m0}}{r} \gg 1,g(l)\right)
\end{eqnarray}
On the r.h.s., $R$ being a short-distance correlator, is ${\cal O}(1)$. Note that 
\begin{eqnarray}
\gamma_{an}(r\ll 2T_{m0})= 1+\frac{\delta}{2}- 2\pi g I_C
\end{eqnarray}

Then,
\begin{eqnarray}
&&\int_{g_0}^{g(l)}\frac{\gamma(g^{\prime})}{\beta(g^{\prime})}dg^{\prime}=\ln{l}
+ \frac{1}{2}\ln\left[\frac{\sinh\left(A\ln{l} + \tanh^{-1}\frac{A}{\delta_0}\right)}{\sinh\left(\tanh^{-1}\frac{A}{\delta_0}\right)}\right]
\nonumber \\
&&- \frac{2\pi I_C A}{B}\ln\left[\frac{\tanh\left(\frac{A}{2}\ln{l} + \frac{1}{2}
\tanh^{-1}\frac{A}{\delta_0}\right)}{\tanh\left(\frac{1}{2}\tanh^{-1}\frac{A}{\delta_0}\right)}\right]
\end{eqnarray}
where $B = \pi \sqrt{c_1}\left(\frac{\gamma^2K_0}{4}\right)\left(\frac{16}{\gamma^2 K_0}-1\right)^{3/4}$.

The correlation function within the light-cone is given by setting $l=\Lambda_0 r$ which gives,
\begin{eqnarray}
&&R\left(\Lambda_0 r, r \ll 2T_{m0}\ll 1/\eta,g_0\right)\simeq \nonumber \\
&&\frac{1}{\left(\Lambda_0 r\right)^{1+A/2}}\sqrt{\frac{1}{1-(\Lambda_0 r)^{-2A}}}
\left[\frac{1-(\Lambda_0 r)^{-A}}{1+(\Lambda_0r)^{-A}}\right]^{\frac{2\pi I_C}{B}}
\end{eqnarray}
For $A\ln{r} \ll 1$, we obtain the following logarithmic correction to scaling,
\begin{eqnarray}
R \sim \frac{1}{r}\left(\ln{r}\right)^{\frac{2\pi I_C}{B}-\frac{1}{2}}
\end{eqnarray}
In the above expression the exponent $\theta = \frac{2\pi I_C}{B}-\frac{1}{2}$ approaches $1/2$ as $K_0$ approaches $K$.

Let us now discuss how $I_C$ behaves for spatial separations on the light-cone. Simplifying the full expression 
for $I_C$ in 
Eqs.~(\ref{ic1}),~(\ref{ic2})
by noting that for $r\gg 1, t\gg 1, r = 2t$,
\begin{eqnarray}
\sqrt{\frac{1+\left(2t-t^{\prime}+x^{\prime}\right)^2}{1+4t^2}}\simeq 1 \nonumber \\ 
\sqrt{\frac{1+\left(2t - t^{\prime}-x^{\prime}\right)^2}{1+4(t-t^{\prime})^2}}  \simeq 1 \nonumber \\
\sqrt{\frac{1+\left(2t-t^{\prime}+x^{\prime}-r\right)^2}{1+4t^2}} = \sqrt{\frac{1+\left(t^{\prime}-x^{\prime}\right)^2}{1+4t^2}}
\nonumber \\
\sqrt{\frac{1+\left(2t-t^{\prime}-x^{\prime}+r\right)^2}{1+4(t-t^{\prime})^2}} \simeq 2 \nonumber 
\end{eqnarray}
one may write,
\begin{widetext}
\begin{eqnarray}
&&I_C(r=2t, r\gg 1, K_{neq}=2) = 2^{-K_{tr}/2}{\rm Lt}_{r \rightarrow \infty}\frac{r^{2+\frac{K_{tr}}{2}}}{2\pi}
\int_{-\infty}^{\infty}dx^{\prime}\int_0^{r/2} dt^{\prime}\nonumber \\
&&\sin\left[\frac{K_{eq}}{2}\biggl\{ \tan^{-1}\left(t^{\prime}+x^{\prime}\right)
+\tan^{-1}\left(t^{\prime}-x^{\prime}\right) 
+ \tan^{-1}\left(t^{\prime}+x^{\prime}-r\right)
+\tan^{-1}\left(t^{\prime}-x^{\prime}+r\right)\biggr\}
\right]\nonumber \\
&&\times \frac{1}{\sqrt{1+\left(t^{\prime}+x^{\prime}\right)^2}}
\left(\frac{1}{\sqrt{1+\left(t^{\prime}-x^{\prime}\right)^2}}\right)^{1+\frac{K_{tr}}{2}}
\frac{1}{\sqrt{1+\left(t^{\prime}+x^{\prime}-r\right)^2}} 
\frac{1}{\sqrt{1+\left(t^{\prime}-x^{\prime}+r\right)^2}}\nonumber \\
&&\times \biggl\{
\frac{1}{1+\left(t^{\prime}+ x^{\prime}\right)^2}
+ \frac{\left(1+\frac{K_{tr}}{2}\right)}{1+\left(t^{\prime}-x^{\prime}\right)^2}
+\frac{1}{1+\left(t^{\prime}+ x^{\prime}-r\right)^2}+ \frac{1}{1+\left(t^{\prime}-x^{\prime}+r\right)^2}
\biggr\}\nonumber \\
&&+ 2^{-K_{tr}/2}{\rm Lt}_{r \rightarrow \infty} \frac{r^{2+\frac{K_{tr}}{2}}}{2\pi}\left(\frac{K_{eq}}{2}\right)
\int_{-\infty}^{\infty}dx^{\prime}\int_0^{r/2}dt^{\prime}\nonumber \\
&&\cos\left[\frac{K_{eq}}{2}\biggl\{ \tan^{-1}\left(t^{\prime}+x^{\prime}\right)
+\tan^{-1}\left(t^{\prime}-x^{\prime}\right) 
 + \tan^{-1}\left(t^{\prime}+x^{\prime}-r\right)
+\tan^{-1}\left(t^{\prime}-x^{\prime}+r\right)\biggr\}\right]\nonumber \\
&&\times \frac{1}{\sqrt{1+\left(t^{\prime}+x^{\prime}\right)^2}}
\left(\frac{1}{\sqrt{1+\left(t^{\prime}-x^{\prime}\right)^2}}\right)^{1+\frac{K_{tr}}{2}}
\frac{1}{\sqrt{1+\left(t^{\prime}+x^{\prime}-r\right)^2}} 
\frac{1}{\sqrt{1+\left(t^{\prime}-x^{\prime}+r\right)^2}}\nonumber \\
&&\times \left[
\frac{\left(t^{\prime}+x^{\prime}\right)}{1+\left(t^{\prime}+ x^{\prime}\right)^2}
+ \frac{\left(t^{\prime}-x^{\prime}\right)}{1+\left(t^{\prime}-x^{\prime}\right)^2}
+\frac{\left(t^{\prime}+x^{\prime}-r\right)}{1+\left(t^{\prime}+ x^{\prime}-r\right)^2}
+ \frac{\left(t^{\prime}-x^{\prime}+r)\right)}{1+\left(t^{\prime}-x^{\prime}+r\right)^2}
\right]
\end{eqnarray}
\end{widetext}
The above implies that 
\begin{eqnarray}
I_C(r=2t, r\gg 1)\sim r^{K_{tr}/2}
\end{eqnarray}
Thus the scaling for spatial separations on the light-cone which we 
found for a pure lattice quench, is lost for a simultaneous lattice and interaction
quench when $K_{tr}>0$ (i.e., $\frac{\gamma^2K_0}{4}> 2$). This
corresponds to a situation where the cosine potential is an irrelevant perturbation in the initial state.
However scaling holds when $K_{tr}\leq 0$ where $I_C$ either approaches zero at large distances or is a
constant. This case of $K_{tr}\leq 0$ corresponds to the cosine potential being a marginal or a relevant perturbation
in the initial state. 

We now turn to the behavior of the correlator for points outside the light-cone. In the previous section
we found that for a pure lattice quench, scaling holds outside the light-cone. We would like to
explore whether this continues to be the case for a simultaneous lattice and interaction quench.
We make the following approximations
in Eqs.~(\ref{ic1}),~(\ref{ic2})
by noting that for $r\gg 1, t\gg 1, r \gg  2t$,
\begin{eqnarray}
\sqrt{\frac{1+\left(2t-t^{\prime}+x^{\prime}\right)^2}{1+4t^2}}\simeq 1 \nonumber \\ 
\sqrt{\frac{1+\left(2t - t^{\prime}-x^{\prime}\right)^2}{1+4(t-t^{\prime})^2}}  \simeq 1 \nonumber \\
\sqrt{\frac{1+\left(2t-t^{\prime}+x^{\prime}-r\right)^2}{1+4t^2}} \simeq \frac{r}{2t}
\nonumber \\
\sqrt{\frac{1+\left(2t-t^{\prime}-x^{\prime}+r\right)^2}{1+4(t-t^{\prime})^2}} \simeq \frac{r}{2t} \nonumber 
\end{eqnarray}
to obtain,
\begin{widetext}
\begin{eqnarray}
&&I_C(r,t \gg 1; r \gg 2 t, K_{neq}=2)\simeq
\frac{r^{2}}{2\pi}\left(\frac{2t}{r}\right)^{K_{tr}}
\int_{-\infty}^{\infty}dx^{\prime}\int_0^{t} dt^{\prime}\nonumber \\
&&\sin\left[\frac{K_{eq}}{2}\biggl\{ \tan^{-1}\left(t^{\prime}+x^{\prime}\right)
+\tan^{-1}\left(t^{\prime}-x^{\prime}\right) 
+ \tan^{-1}\left(t^{\prime}+x^{\prime}-r\right)
+\tan^{-1}\left(t^{\prime}-x^{\prime}+r\right)\biggr\}
\right]\nonumber \\
&&\times \frac{1}{\sqrt{1+\left(t^{\prime}+x^{\prime}\right)^2}}
\frac{1}{\sqrt{1+\left(t^{\prime}-x^{\prime}\right)^2}}
\frac{1}{\sqrt{1+\left(t^{\prime}+x^{\prime}-r\right)^2}} 
\frac{1}{\sqrt{1+\left(t^{\prime}-x^{\prime}+r\right)^2}}\nonumber \\
&&\times \biggl\{
\frac{1}{1+\left(t^{\prime}+ x^{\prime}\right)^2}
+ \frac{1}{1+\left(t^{\prime}-x^{\prime}\right)^2}
+\frac{1}{1+\left(t^{\prime}+ x^{\prime}-r\right)^2}+ \frac{1}{1+\left(t^{\prime}-x^{\prime}+r\right)^2}
\biggr\}\nonumber \\
&&+\frac{r^{2}}{2\pi}\left(\frac{2t}{r}\right)^{K_{tr}} \left(\frac{K_{eq}}{2}\right)
\int_{-\infty}^{\infty}dx^{\prime}\int_0^{t}dt^{\prime}\nonumber \\
&&\cos\left[\frac{K_{eq}}{2}\biggl\{ \tan^{-1}\left(t^{\prime}+x^{\prime}\right)
+\tan^{-1}\left(t^{\prime}-x^{\prime}\right) 
 + \tan^{-1}\left(t^{\prime}+x^{\prime}-r\right)
+\tan^{-1}\left(t^{\prime}-x^{\prime}+r\right)\biggr\}\right]\nonumber \\
&&\times \frac{1}{\sqrt{1+\left(t^{\prime}+x^{\prime}\right)^2}}
\frac{1}{\sqrt{1+\left(t^{\prime}-x^{\prime}\right)^2}}
\frac{1}{\sqrt{1+\left(t^{\prime}+x^{\prime}-r\right)^2}} 
\frac{1}{\sqrt{1+\left(t^{\prime}-x^{\prime}+r\right)^2}}\nonumber \\
&&\times \left[
\frac{\left(t^{\prime}+x^{\prime}\right)}{1+\left(t^{\prime}+ x^{\prime}\right)^2}
+ \frac{\left(t^{\prime}-x^{\prime}\right)}{1+\left(t^{\prime}-x^{\prime}\right)^2}
+\frac{\left(t^{\prime}+x^{\prime}-r\right)}{1+\left(t^{\prime}+ x^{\prime}-r\right)^2}
+ \frac{\left(t^{\prime}-x^{\prime}+r)\right)}{1+\left(t^{\prime}-x^{\prime}+r\right)^2}
\right]
\end{eqnarray}
\end{widetext}

The above implies that outside the light-cone,
\begin{eqnarray}
I_C\left(r\gg 2t, r\gg 1, t\gg 1 \right)\sim \left(\frac{2t}{r}\right)^{K_{tr}}
\end{eqnarray}
Thus outside the light-cone, we find that if $K_{tr}> 0$, {\sl i.e.}, the 
cosine potential is an irrelevant perturbation in the initial state, then $I_C$ goes to zero with distance 
as a power-law. However if $K_{tr}<0$, then $I_C$ grows with distance, and the perturbative correction to the
correlation function becomes large. This result is expected as the cosine potential for $K_{tr}<0$ 
is a relevant perturbation in the initial state. 
Thus perturbation theory at initial times outside the light-cone is violated. An
example of this was also discussed in Section~\ref{pertR}, and shown in Fig.~\ref{figR1}.

\section{Summary and Conclusions} \label{summary}

We have studied quench dynamics in a generic strongly correlated one dimensional system which is represented by the
quantum sine-Gordon model. We develop a novel time-dependent renormalization group approach which reveals that the
dynamics after a quantum quench can be quite rich by being characterized by several time-scales (Fig.~\ref{fig3}). One is a 
perturbatively accessible short time scale ($T_m \ll 1/\Lambda$), the second is an intermediate time prethermalized regime where
inelastic effects are small ($1\ll T_m \ll 1/\eta$) and the system can show
universal scaling behavior, and the third is a longer time scale ($T_m\gg 1/\eta$)  where inelastic scattering
is strong, leading to thermal behavior. In this paper we explicitly derived 
a CS-like differential equation~(\ref{cz}) for a two-point correlation function, and solved it in the prethermalized regime.
This CS equation shows that even in the universal prethermalized regime, three distinctly different scaling regimes 
can exist that are summarized in Fig.~\ref{fig2}. One is for spatial separations of the local operators outside the light-cone,
the other is for spatial separations on the light-cone and the third is for spatial separations inside the light-cone.

When only the cosine potential is quenched, and for a final Hamiltonian $H_f$ which is 
at the critical point, the results for the correlator in the
three scaling regimes are given in equations~(\ref{czsol2}),~(\ref{czsol3}) and ~(\ref{czsol4}). Universal logarithmic corrections 
due to the marginal cosine potential are found, and  consistent with the horizon effect, the correlator 
is most enhanced right on the light-cone. 
For a final Hamiltonian with parameters that
correspond to slight deviations away from the critical point, the result for the correlation function outside the light-cone
is given in Eq.~(\ref{Rolca}), and inside the light-cone is given in Eq.~(\ref{Rilca}). 

For more complicated quenches where the initial Luttinger parameter is quenched at the same time as the cosine
potential is switched on, scaling holds within the light-cone, and the results are given in Eqs.~(\ref{Rgen}) and ~(\ref{log3}).
Whether scaling holds on the light-cone and outside it depends on the initial state (or the initial Luttinger
parameter). In particular when the cosine potential is a relevant or marginal perturbation in the initial state, 
scaling holds on the light-cone, otherwise it is violated. In contrast outside the light-cone, if the cosine
potential is a relevant perturbation in the initial state, the perturbative corrections grow with distance. The latter behavior
is consistent with our understanding that outside the light-cone the behavior is primarily determined by the initial state. Thus
even though the cosine potential may be a marginal or irrelevant perturbation for the final Hamiltonian (or the wave-function
at long times after the quench), if it is a relevant
perturbation for the initial Hamiltonian, perturbation theory will be violated outside the light-cone.
 
There are many interesting open questions. One is to generalize the results of this paper to unequal time correlation
functions with the aim of studying issues such as aging~\cite{Gambassi05,Biroli12}. 
The current paper focuses on the prethermalized regime where 
the main assumption is that inelastic effects being weak, the nonequilibrium boson distribution function
generated due to the quench hardly changes in time. 
The dynamics in the thermal regime, and in particular how the boson
distribution function evolves in time due to strong inelastic scattering is
also very interesting to study. If the time-scale $1/\eta$ is very short ({\sl i.e.}, the quench amplitude is large),
so that the prethermalized regime is almost absent, a 
quantum kinetic equation may be employed to study how the boson distribution function evolves in time~\cite{Tavora13}. 
When $1/\eta$ is large resulting in a long prethermalized regime, understanding the difficult problem of how
the crossover in time from the prethermalized to the thermalized regime occurs, and 
observing this in numerical studies~\cite{Karrasch12} is also an important open question.
It is also interesting to study the regime where the cosine potential is a relevant perturbation either in the initial
or final state or both. When the cosine potential is irrelevant in the initial state, but relevant in the final state
after the quench, RG may be used to identify a critical time after the quench when
perturbation theory breaks down~\cite{Mitra12b}.  
Studying the time-evolution of two-point correlation functions
and also quantities such as the Loschmidt echo when the cosine potential is relevant are important open questions.
Finally an interesting direction to pursue is to employ the approach of this paper to
study quench dynamics in quantum field theories in higher spatial dimensions.

{\sl Acknowledgments:} The author is deeply indebted to Fabian Essler for many helpful discussions and for a critical
reading of the manuscript. This research was supported by the National Science Foundation under 
Grant No. PHY11-25915 and Grant No. DMR-1004589.

\begin{appendix}

\section{Expressions for $I_{u,K,\eta,T_{eff}}$}

\begin{eqnarray}
&&I_{T_{eff}}(T_m\Lambda)= \int_{-\infty}^{\infty} d\bar{r} \int_{-2T_m\Lambda}^{2T_m\Lambda} d\bar{\tau}\nonumber \\
&&{\rm Re}\left[B(\bar{r},T_m\Lambda,\bar{\tau})\right]\\
&&I_{\eta}(T_m\Lambda)= -\int_{-\infty}^{\infty} d\bar{r} \int_{-2T_m\Lambda}^{2T_m\Lambda} d\bar{\tau}\bar{\tau}  \nonumber \\
&&{\rm Im}\left[B(\bar{r},T_m\Lambda,\bar{\tau})\right]\\
&&I_{u}(T_m\Lambda)= -\int_{-\infty}^{\infty} d\bar{r} \int_{0}^{2T_m\Lambda } d\bar{\tau} 
\left(\bar{r}^2+\bar{\tau}^2\right)\nonumber\\
&&{\rm Im}\left[B(\bar{r},T_m\Lambda,\bar{\tau})\right]\\
&&I_{K}(T_m\Lambda)= -\int_{-\infty}^{\infty} d\bar{r} \int_{0}^{2T_m\Lambda} d\bar{\tau}\left(\bar{r}^2- 
\bar{\tau}^2\right)\nonumber \\
&&{\rm Im}\left[B(\bar{r},T_m\Lambda ,\bar{\tau})\right]\label{IK}
\end{eqnarray}
where ${\rm Re}[B]=(B+B^*)/2$, ${\rm Im}[B]$=$(B-B^*)/(2i)$, and
\begin{eqnarray}
B(\bar{r},T_m\Lambda,\bar{\tau})=C_{+-,1}(\bar{r},T_m\Lambda,\bar{\tau})F(\bar{r},T_m\Lambda,\bar{\tau})
\end{eqnarray}
with $C_{+-,1}(\bar{r},T_m\Lambda,\bar{\tau})=\langle e^{i\phi_+(\bar{r},\bar{\tau}+T_m\Lambda/2)}e^{-i\phi_-(0,\bar{\tau}-T_m\Lambda/2)} \rangle$. 
This quantity within leading order in perturbation theory is,
\begin{eqnarray}
&&C_{+-,1}(\bar{r},T_m\Lambda,\bar{\tau})
=\left[\frac{1}{\sqrt{1 + (\bar{\tau}+\bar{r})^2}}
\frac{1}{\sqrt{1 + (\bar{\tau}-\bar{r})^2}}\right]^{K_{neq}}
\nonumber \\
&&
\left[\frac{\sqrt{1 + \{2(T_m\Lambda+\bar{\tau}/2)\}^2}}{\sqrt{1 + (2T_m\Lambda+\bar{r})^2}}\frac{\sqrt{1
+ \{2 (T_m\Lambda-\bar{\tau}/2)\}^2}}{\sqrt{1 + (2T_m\Lambda-\bar{r})^2}}\right]^{K_{tr}}\nonumber \\
&&\times e^{-i K_{eq}\left[
\tan^{-1}\left(\bar{\tau}+{\bar r}\right)
+ \tan^{-1}\left(\bar{\tau}-\bar{r}\right)
\right]}
\end{eqnarray}
while $F$ is given by
\begin{eqnarray}
&&F(\bar{r},T_m\Lambda,\bar{\tau}) = K_{neq}\left[\frac{1}{1 + \left(\bar{\tau}+\bar{r}\right)^2} +
\frac{1}{1 + \left(\bar{\tau}-\bar{r}\right)^2}\right]\nonumber \\
&&+K_{tr}\left[\frac{1}{1 + \left(2T_m\Lambda +\bar{r}\right)^2} + \frac{1}{1 +
\left(2T_m\Lambda-\bar{r}\right)^2}\right]\nonumber \\
&&- i K_{eq}\left[\frac{\bar{\tau}+\bar{r}}{1 + \left(\bar{\tau}+\bar{r}\right)^2}
+ \frac{\bar{\tau}-\bar{r}}{1 + \left(\bar{\tau}-\bar{r}\right)^2}\right]
\end{eqnarray}

\end{appendix}

\begin{thebibliography}{58}
\bibitem [{com()}]{comm1}%
  \BibitemOpen
  \href@noop {} {}\bibinfo {note} {Here we have dropped the $\langle \nabla
  \phi(1) \nabla \phi(2)\rangle$ term as it decays faster in position, and is
  therefore less important than $\langle
  \cos\{\gamma\phi(1)\}\cos\{\gamma\phi(2)\}\rangle$ near the critical point
  where $V_{sg}$ is marginal. This holds true both in and out of
  equilibrium.}\BibitemShut {Stop}%
\end{thebibliography}
%

\end{document}